# Phenotypic Trait of Particle Geometries

Seung Jae Lee[1,*], Moochul Shin[2], Chang Hoon Lee[2] and Priya Tripathi[1]

[1]Department of Civil & Environmental Engineering, Florida International University, Miami, FL, USA
[2]Department of Civil & Environmental Engineering, Western New England University, Springfield, MA, USA

## Abstract

People of a race appear different but share a 'phenotypic trait' due to a common genetic origin. Mineral particles are like humans: they appear different despite having a same geological origin. Then, do the particles have some sort of 'phenotypic trait' in the geometries as we do? How can we characterize the phenotypic trait of particle geometries? This paper discusses a new perspective on how the phenotypic trait can be discovered in the particle geometries and how the 'variation' and 'average' of the geometry can be quantified. The key idea is using the power-law between particle surface-area-to-volume ratio (A/V) and the particle volume (V) that uncovers the phenotypic trait in terms of $\alpha$ and $\beta^*$: From the log-transformed relation of $V = (A/V)^\alpha \times \beta$, the power value $\alpha$ represents the relation between shape and size, while the term $\beta^*$ (evaluated by fixing $\alpha = -3$) informs the angularity of the average shape in the granular material. In other words, $\alpha$ represents the 'variation' of the geometry while $\beta^*$ is concerned with the 'average' geometry of a granular material. Furthermore, this study finds that A/V and V can be also used to characterize individual particle shape in terms of Wadell's true Sphericity ($S$). This paper also revisits the $M$ = A/V×L/6 concept originally introduced by Su et al. [1] and finds the shape index $M$ is an extended form of $S$ providing additional information about the particle elongation. Therefore, the proposed method using A/V and V provides a unified approach that can characterize the particle geometry at multiple scales from granular material to a single particle.

---

[*] Corresponding author: sjlee@fiu.edu

# 1 INTRODUCTION

The particle geometry (i.e., shape, volume, surface area, and size) is fundamental information to understand the physical properties and behavior of granular material [2–6]. Typically, mineral particles usually come in a variety of shapes and sizes. In a sense, mineral particles are like humans: people of a race do not look alike and have different appearances despite a common genetic origin. Likewise, mineral particles look different even if they have same geological origin. However, people share a 'phenotypic trait' that may be found from the 'average face' of each race [7, 8] as in the examples of Figure 1. Then, we can ask some questions: Can we find such a phenotypic trait in the particles in terms of the 'average' geometry? Also, how can we systematically quantify the geometry 'variation' in the particles having a common geological origin? To address these questions, this paper proposes a new approach using a power-law between particle surface-area-to-volume ratio (A/V) and the particle volume (V) that uncovers the phenotypic trait (i.e., average and variation) of particle geometries of a granular material. In addition, this paper demonstrates Wadell's true Sphericity ($S$) that characterizes the individual particle shape can be also quantified in terms of A/V and V. Therefore, A/V and V can be used consistently to describe the particle geometry at multiple scales from granular material to single particle. Furthermore, this paper discusses $M$ = A/V×L/6 concept introduced by Su et al. [1] is an extended form of $S$ that provides the additional information about the particle elongation. Section 2 discusses the use of the power law between A/V and V to identify the phenotypic trait of granular material. Section 3 discusses the use of individual A/V and V data to identify the individual particle shape.

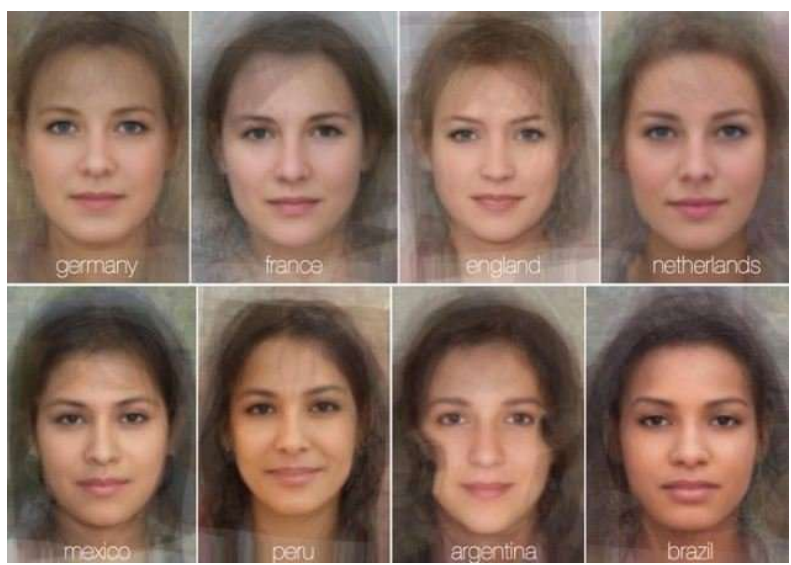

Figure 1. Average face of women in some countries (Image used under CC-BY-4.0 license [8])

## 2  PHENOTYPIC TRAIT OF GRANULAR MATERIAL

The particle surface-area-to-volume ratio (A/V) is the key information to characterize the particle shape [1, 9], and the relation between A/V and particle volume (V) of a granular material (i.e., a group of particles) can be approximated by a 'power-law' [10]. We find this power-law can be used to uncover the 'phenotypic trait' of the particle geometries in a granular material.

Particle volume (V) is a 3-dimensional measure and particle surface area (A) is a 2-dimensional measure. Therefore, the relation between A and V can be formulated as shown in Equation (1), where λ is a constant that depends on particle geometry. Equation (1) can be reformulated to Equation (2) in terms of A/V and V, then can be generalized to Equation (3) with α and β. This power-law relation between A/V and V is presented as a linear plot in a log-log space as presented in Equation (4)[†]. Therefore, the power value α represents the *slope* of the log-transformed plot, and the term β represents the plot's *intercept* at A/V = 1.

$$V = A^{3/2} \times \lambda \tag{1}$$

$$V = (A/V)^{-3} \times 1/\lambda^2 \tag{2}$$

$$V = (A/V)^{\alpha} \times \beta \tag{3}$$

$$\log(V) = \alpha \times \log(A/V) + \log(\beta) \tag{4}$$

### 2.1  Particles of an Identical Shape

As shown in Equation (2), the power value α is -3 for a particle. This implies α (slope of the log-transformed plot) is -3 if all particles in a granular material have a same shape. Figure 2 shows an example in which the A/V and V values of three particle groups are plotted in a log-log space. Each group has an identical shape: Group 1 is composed of 15 spheres, Group 2 has 15 cubes, and Group 3 has 15 regular tetrahedrons. The particles in each group have various sizes of about 0.3 ~10 mm measured in terms of D. The size D is the diameter of reference sphere having the same volume as the particle, which can be computed as shown in Equation (5).

---

[†] Throughout this paper, the common logarithm (base 10) is considered, and we omit writing of the base 10 in the logarithm notation.

$$D = 2 \times (3V / (4\pi))^{1/3} \qquad (5)$$

As shown in Figure 2, the A/V and V relation can be approximated with a power function. The slope α is invariably -3 for every group with $R^2 = 1$. There is only one shape per group, so α = -3 indicates the shape does not change with size. (In Section 2.2, we will discuss in detail how α varies as the shape changes with size.) The intercept β is the volume V at A/V = 1 and in this example increases with the angularity of particle geometry. The β is 113.09 (= 36π) for spherical particles, 216 for cubes and 374.12 for tetrahedra with the slope α = -3. The geometric properties of all particles are presented in Table A.1 to Table A.3 in the Appendix. Hereafter, β* is used to indicate a specific β evaluated with a fixed value of α = -3 as shown in Equation (6). The β* is related to the average shape in case particles have different shapes. This will be further discussed in Section 2.2.

$$\log(V) = -3 \times \log(A/V) + \log(\beta^*) \qquad (6)$$

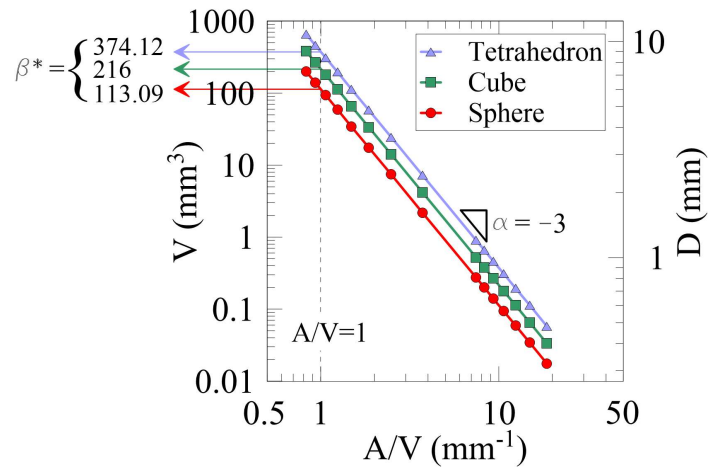

Figure 2. Power-law relation for the three groups of particles with identical shape

As a side note, readers may wonder why Equation (2) is used (in terms of A/V and V) instead of Equation (1) although Equation (1) can be also transformed to a log-log function between A and V. It is certainly possible, but with Equation (1) the data points tend to be located closer and may be difficult to tell apart. For example, Figure 3 shows the power-law relation between A and V where the data points are less distinct compared to Figure 2. However, use of either Equation (1) or (2) seems to be a matter of preference. Note that a different interpretation should be required when using Equation (1). For example, the slope of the log-transformed equation changes to positive (i.e., 3/2). A study using Equation (1) is suggested for future research.

Lee et al. (2021)                              Page 4 of 30

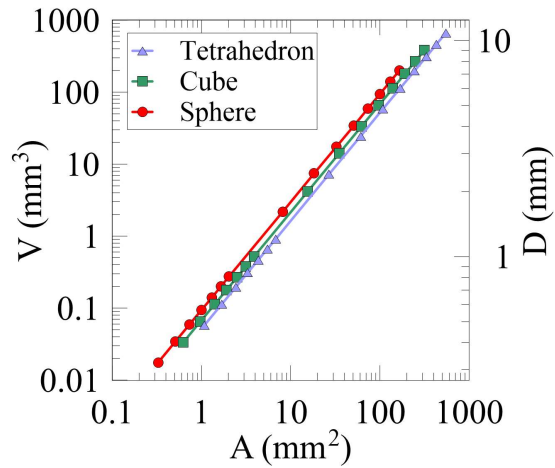

Figure 3. Power-law relation in terms of A and V

## 2.2 Particles with Different Shapes

The next question is how α and β* would change when particles have different shapes within a group? Group A and B are created as the example particle groups with different shapes. Each group consists of three particles (a regular tetrahedron, a cube, and a sphere) selected from the three-shape groups presented in Section 2.1. The difference between Group A and B is the relation between shape and size: in Group A, the more angular shape is larger, (i.e., tetrahedron is the largest, cube is the next, and sphere is the smallest). On the other hand, it is the opposite in Group B (i.e., sphere is the largest, cube is the next, and tetrahedron is the smallest). In Figure 4a and b, the size of three particles is schematically shown and the corresponding data points are highlighted with different colors. The geometric properties of particles in Group A and B are presented in Table A.4.

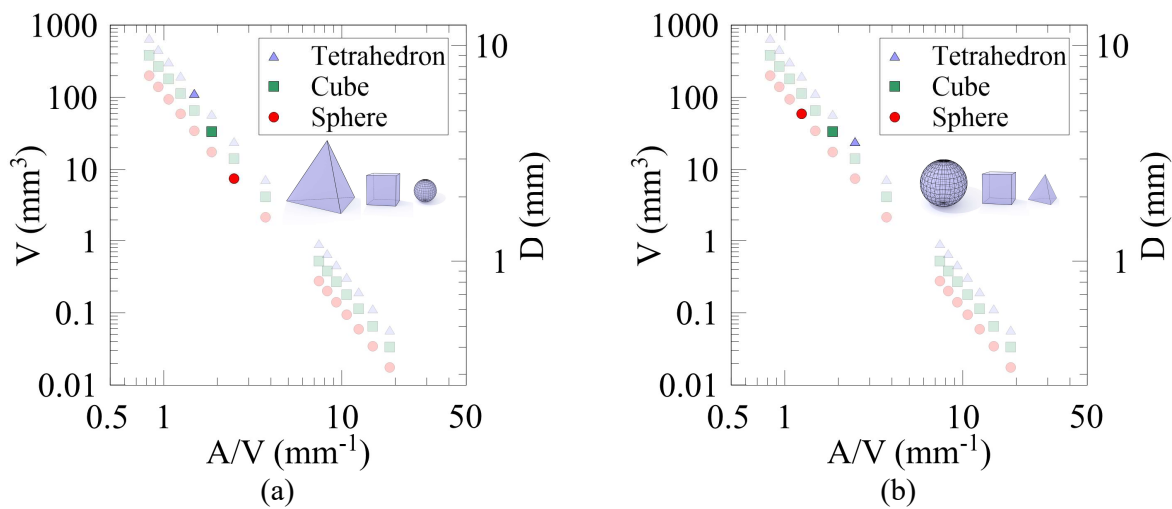

Figure 4. Two groups of particles with three different shapes: (a) Group A – the more angular shape is larger; (b) Group B – the shape-size relation is the opposite to Group A

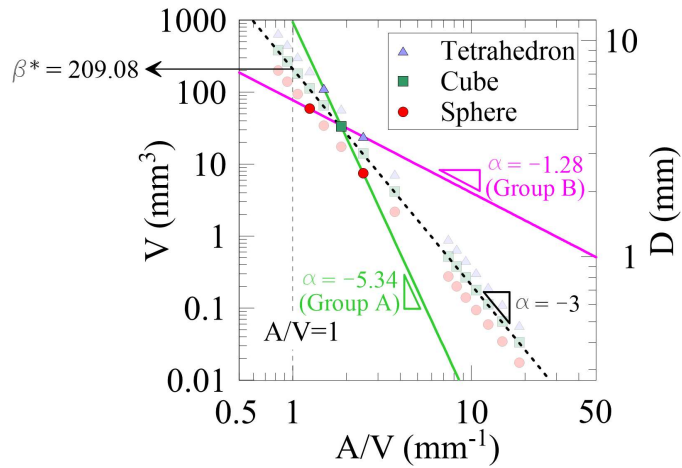

Figure 5. Power-law relations for the two groups of particles with different shapes

### 2.2.1 α – Indication of Variation

Figure 5 shows the power regressions for both groups. Here we can see that the power value α (slope) clearly indicates the relation between shape and size. The power value α is not -3 for both groups. As shown for Group A (α = -5.34), when α is less than -3, the slope is steeper ($|α| > 3$), implying that the larger particles are more angular and smaller particles are more spherical and rounder. As in Group B (α = -1.28), when α is greater than -3, the slope is gentle ($|α| < 3$), indicating that larger particles are more spherical and rounder. On the other hand, α = -3 indicates there is no particular relation between the shape and size. For example, α is -3 if all 45 particles are considered (the dashed line in Figure 5) because the constituent particle shapes do not change with size, i.e., tetrahedron, cube, and sphere. This may be considered as a generalization of the case in Section 2.1 (particles of an identical shape), i.e., if all particles have a same shape, there is no shape-size relation because the shape does not change with size.

### 2.2.2 β* – Indication of Average

The β* is concerned with the average of particle geometries and has three properties as below. The β* is a specific β evaluated with a fixed value of α = -3. Therefore, β ≠ β* if α ≠ -3. For example, the power regression of Group A can be expressed as $V = (A/V)^{-5.34} \times 939.14$, thus β is 939.14 while β* is 209.08 per Equation (6).

First, β* represents the angularity of the average shape. The β* obtained for all 45 particles is 209.08 (which can be computed with the averages of log(V) and log(A/V) as presented in Table A.1 to Table A.3, i.e., the average log(V) of all 45 particles is ~ 0.573 and the average log(A/V) is ~ 0.582, from which β* is 209.08 per Equation (6)). This β* informs the angularity of the average particle shape is close to that of the cube, considering β* of the cubes-only group is 216 (as shown in Figure 2). This makes sense because (i) the average shape is somewhere halfway between the sphere and the tetrahedron, and (ii) there are equal amounts of sphere, cube and tetrahedron within each group (i.e., 15 particles each). For the same reason, the β* evaluated for each Group A and B is also equal to 209.08 because there are equal amounts of sphere, cube and tetrahedron within each group (i.e., one particle each). Figure 6 shows other examples of how β* changes depending on the shape of the constituent particles. The details of the geometric properties are presented in Table A.5. Figure 6a shows β* = 209.08 as in Groups A and B. With an elongated tetrahedron as in Figure 6b, β* increases to 231.31. As the elongated tetrahedron has the higher angularity than the regular tetrahedron in Figure 6a, the average angularity gets higher which is reflected in the higher β*. To the contrary, if there are two spheres in the group (Figure 6c), the β* drops to 179.31, because the additional sphere reduces the average angularity. The β* decreases even further when there are more spheres: With 100 spheres as in Figure 6d, β* is 115.16. The spheres are dominant in the granular material, so β* is close to 113.09 of the spheres-only group (Figure 2). As the number of spheres increases, β* approaches 113.09 asymptotically.

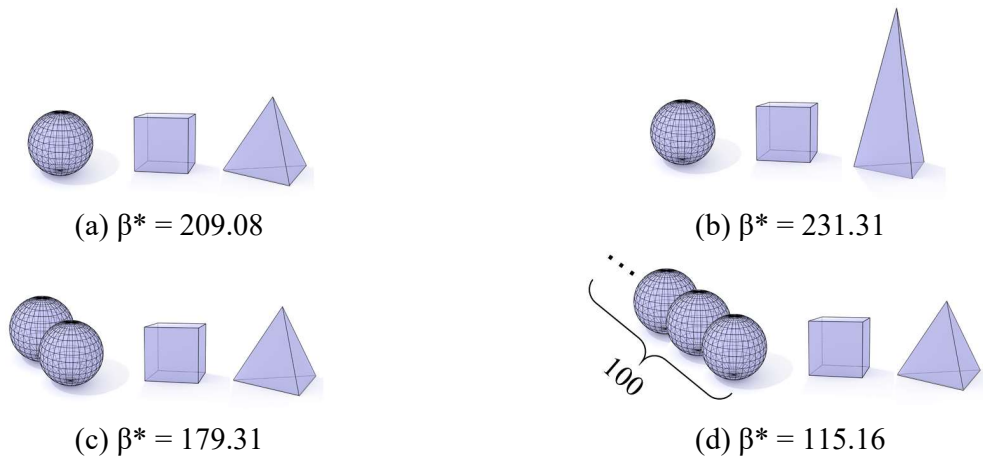

(a) β* = 209.08  (b) β* = 231.31
(c) β* = 179.31  (d) β* = 115.16

Figure 6. Change of β* with different constituent shapes; The length of three edges of the elongated tetrahedron in (b) is 3.2 each, and the length of the other edges is 1.5.

Second, β* is scale-free in the sense that β* is not influenced by the particle size. As shown in Figure 2, α is -3 if particles of different sizes have the same shape. This implies a change in particle size simply moves the data point along the slope α = -3, which does not change the distance between the data point and a line with slope α = -3. Since β* is evaluated with a fixed value of α = -3, the evaluation of β* is not influenced by the particle size. Therefore, Group A and B having the same set of particle shapes show the same β*= 209.08 despite the different shape-size relations. Figure 6a is a variation of Group A and B with the same set of particle shapes. The only difference is that particles in Figure 6a have a unit volume (see Table A.5). Likewise, the same β* is obtained, and the different size has no effect on β*. For another example, a variation of Figure 6b is created with arbitrary sizes as below, and the particle properties are shown in Table 1. The same β* = 231.31 as in Figure 6b is also obtained.

Table 1. Geometric properties of a variation to Figure 6b

| Figure 6b | D (mm) | V (mm$^3$) | A (mm$^2$) | A/V (mm$^{-1}$) | log(V) | log(A/V) | |
|---|---|---|---|---|---|---|---|
| Sphere | 11.664 | 830.985 | 427.444 | 0.514 | 2.920 | -0.289 | |
| Cube | 5.309 | 78.362 | 109.872 | 1.402 | 1.894 | 0.147 | |
| ET* | 6.204 | 125.000 | 199.293 | 1.594 | 2.097 | 0.203 | |
| Average | | | | | 2.304 | 0.020 | β* = 231.31 |

*ET: Elongated Tetrahedron (the length of three edges of the elongated tetrahedron is 16 each, and the length of the other edges is 7.5.)

<u>Third, β* is independent of unit.</u> For the same reason that β* is scale-free as above, a change of unit simply moves the data point along the slope α = -3. Therefore, β* is not affected by the unit. For example, Figure 7 shows the Group B data in both mm and cm, which results in the same β* = 209.08. The data are provided in Table A.4.

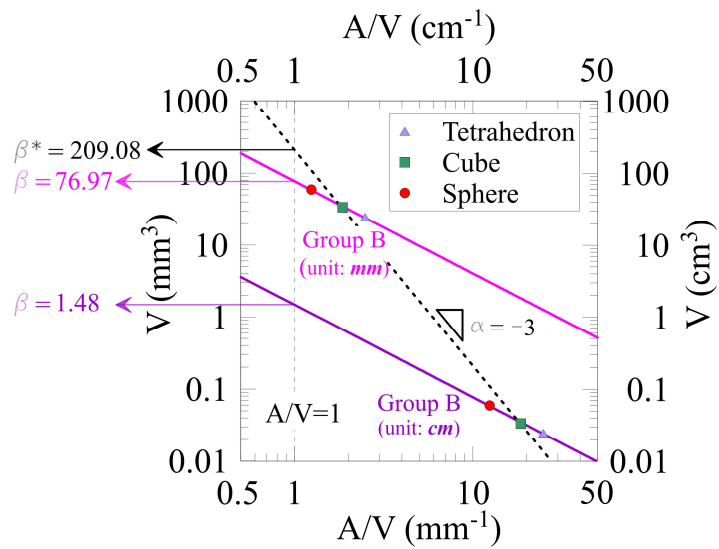

Figure 7. Group B data in millimeter (mm) and centimeter (cm); β* is not influenced by the unit.

It is worth noting that β does not provide information related to the angularity of average shape unlike β*. The intercept β may change with the slope α as shown in Figure 5. Although the constituent shapes in Group A and B are identical, the different shape-size relations (thus different α) change the intercept β. That is, β is affected by the relative size of each shape. For example, in Figure 5, the power regression of Group B is V = (A/V)$^{-1.28}$ × 76.97. Despite Group B has angular shapes in addition to the sphere, β of Group B (= 76.97) is even lower than that of the spheres-only group (β = 113.09) in Figure 2. Furthermore, β changes with the unit unlike β*, which does not make β as a useful index. For example, if *cm* is used (instead of *mm*), the power regression of Group B is expressed as V = (A/V)$^{-1.28}$ × 1.48 as shown in Figure 7.

## 2.3 Examples

We demonstrate the evaluation of α and β* for larger datasets. Section 2.3.1 analyzes a set of 200 polyhedral particles artificially developed in various shapes and sizes. This set includes near-spherical particles which are expected to exhibit a low β* close to that of the spheres (= 113.09). Section 2.3.2 analyzes a set of mineral particles after 3D scanning to see how α and β* are realized for the real particles. Section 2.3.3 analyzes fragmented particles which are expected to exhibit the highest β* value among the three examples due to the presence of angular shapes generated by the breakage.

### 2.3.1 Polyhedral Particles

A larger dataset of 200 polyhedral particles having various shapes and sizes was artificially generated by Su et al. [1]. The particle images are shown in Figure 8 and the geometry information are summarized in Table A.6. This set of particles is composed of two groups: (i) near-spherical (particle ID: 1-100) and (ii) mixed shapes (particle ID: 101-200). When creating the group of mixed shapes, a specific shape-size relation was considered such that the smaller particles tend to have more non-spherical shapes. Figure 9 shows the phenotypic traits of particle geometries from the A/V and V relation. As shown in the figure, two different traits are clearly seen, indicating two different sets are mixed in the 200 particles. For one group of the data, the evaluated α is -2.99 and β* is 116.29, and for the other group, α is -2.64 and β* is 155.69. The first power regression with α close to -3 indicates that all particle shapes are identical, and β* = 116.29 informs the average shape is close to sphere (given that β* of the sphere-only group is 113.09). Therefore, this set of α and β* correctly pinpoints the near-spherical particle group. On the other hand, |α| of the other regression has a gentler slope of 2.64, which indicates that the smaller particles have more irregular shapes other than sphere. The overall angularity is higher due to the mixed shapes, which is realized with the higher β* (= 155.69) than the near-spherical particle group.

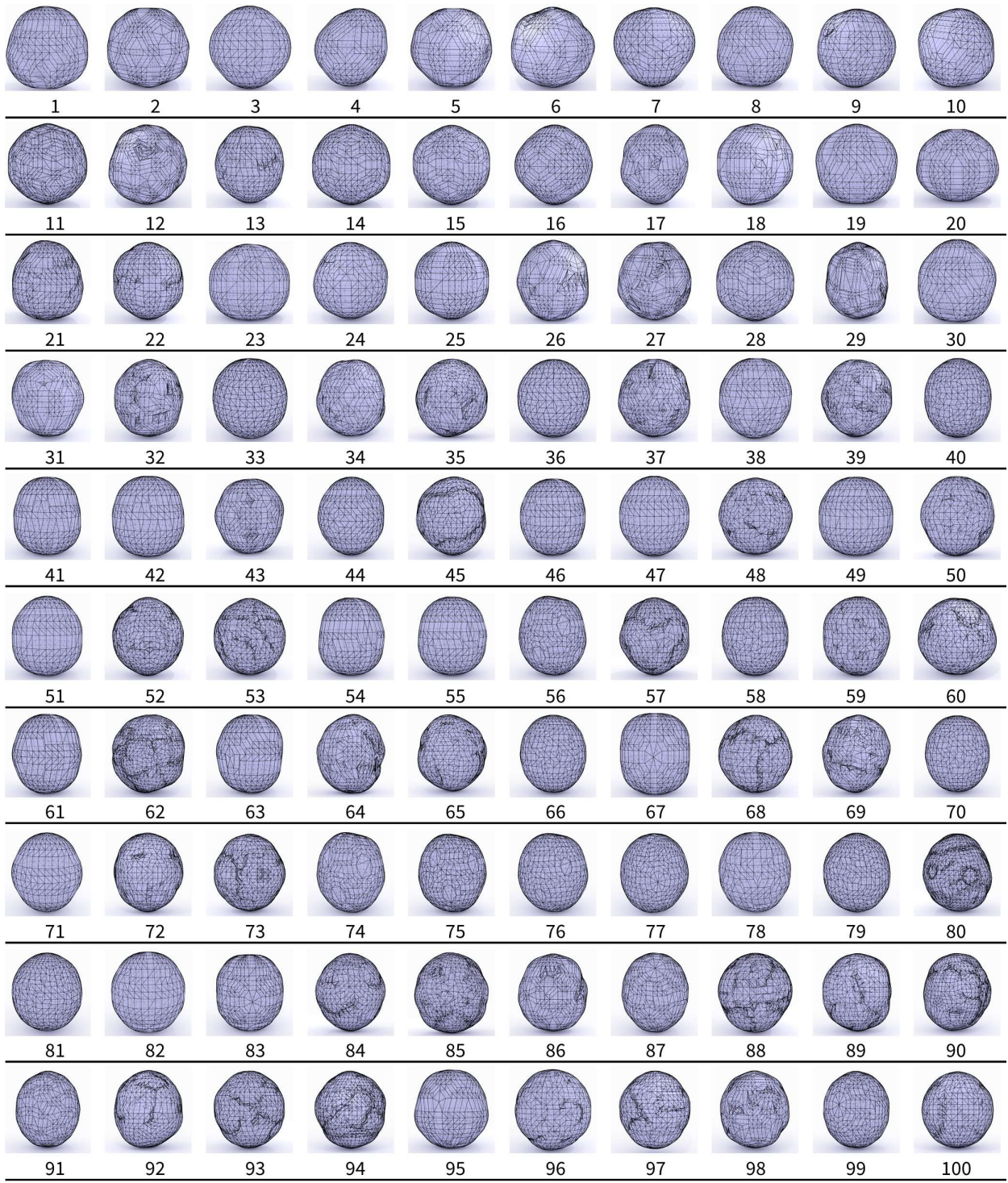

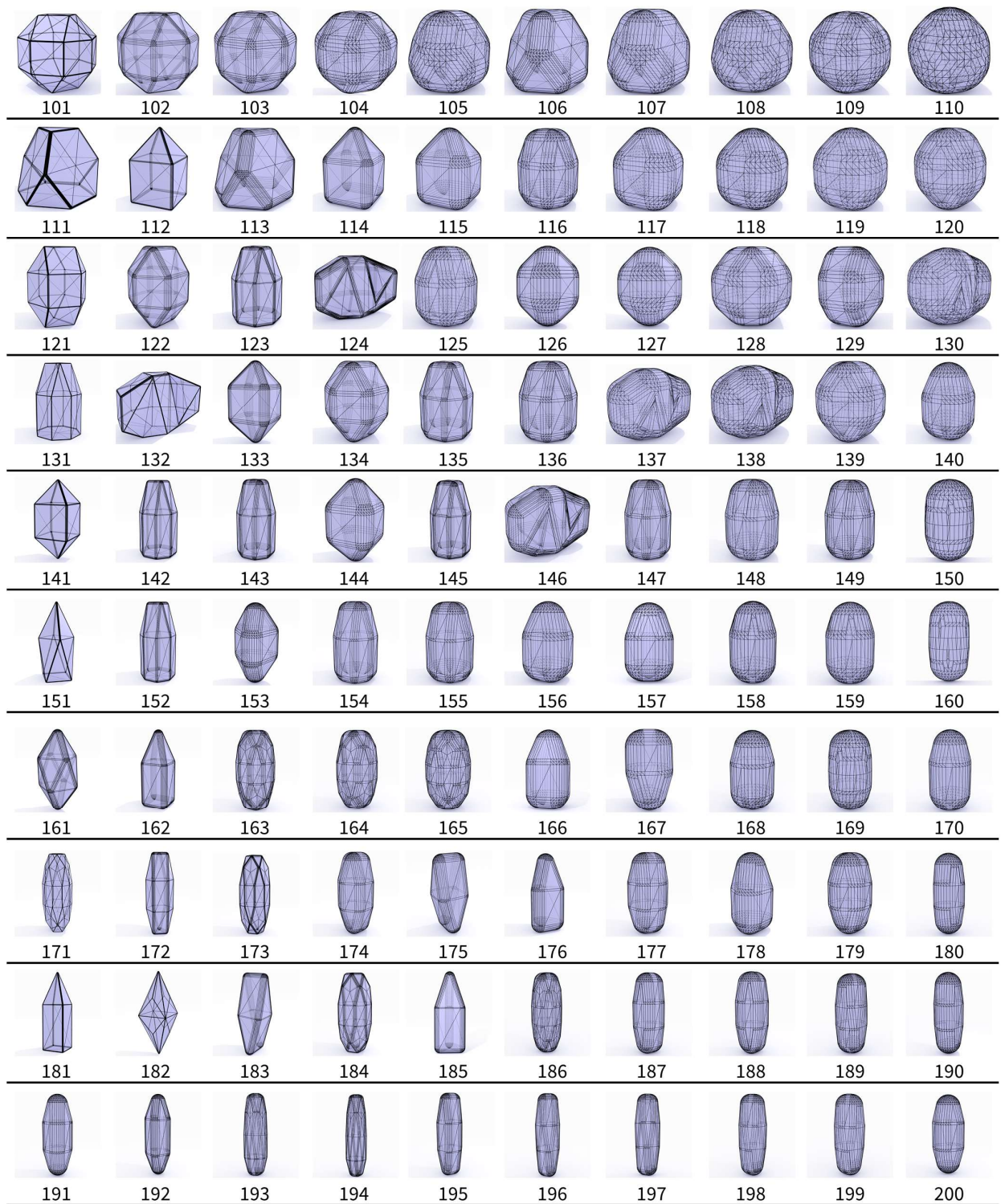

Figure 8. Artificially generated 200 polyhedral particles (image reproduced from Su et al. [1])

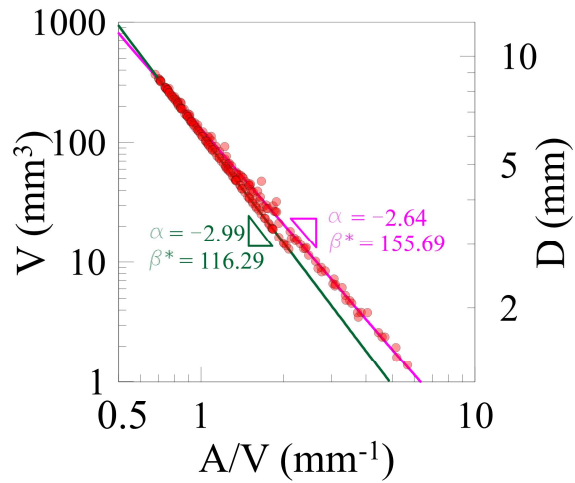

Figure 9. Two different phenotypic traits in the 200 particles

2.3.2 Mineral Particles

A set of 60 limestone particles provided by the Florida Department of Transportation are analyzed for the evaluation of α and β*. Figure 10 shows the limestone particles. The A/V and V data are obtained through 3D scanning. This study adopts the Polyga C504 structured light scanner capable of obtaining high-quality 3D images [11]. The scanner is equipped with a pair of high-resolution cameras that can capture up to 6-micron details. The companion Polyga software creates the 3D images in a standard 3D image format including OBJ and STL, from which the A/V and V data is numerically obtained. The geometric properties are presented in Table A.7. Figure 11 shows the phenotypic traits of particle geometries from the power law relation. The obtained α is -2.81 and β* is 227.55. The α value indicates that the larger particles tend to be slightly more spherical than smaller particles. Compared to the 200 polyhedral particles in Section 2.3.1, the β* is higher due to the overall higher angularity of the mineral particles as shown in Figure 10.

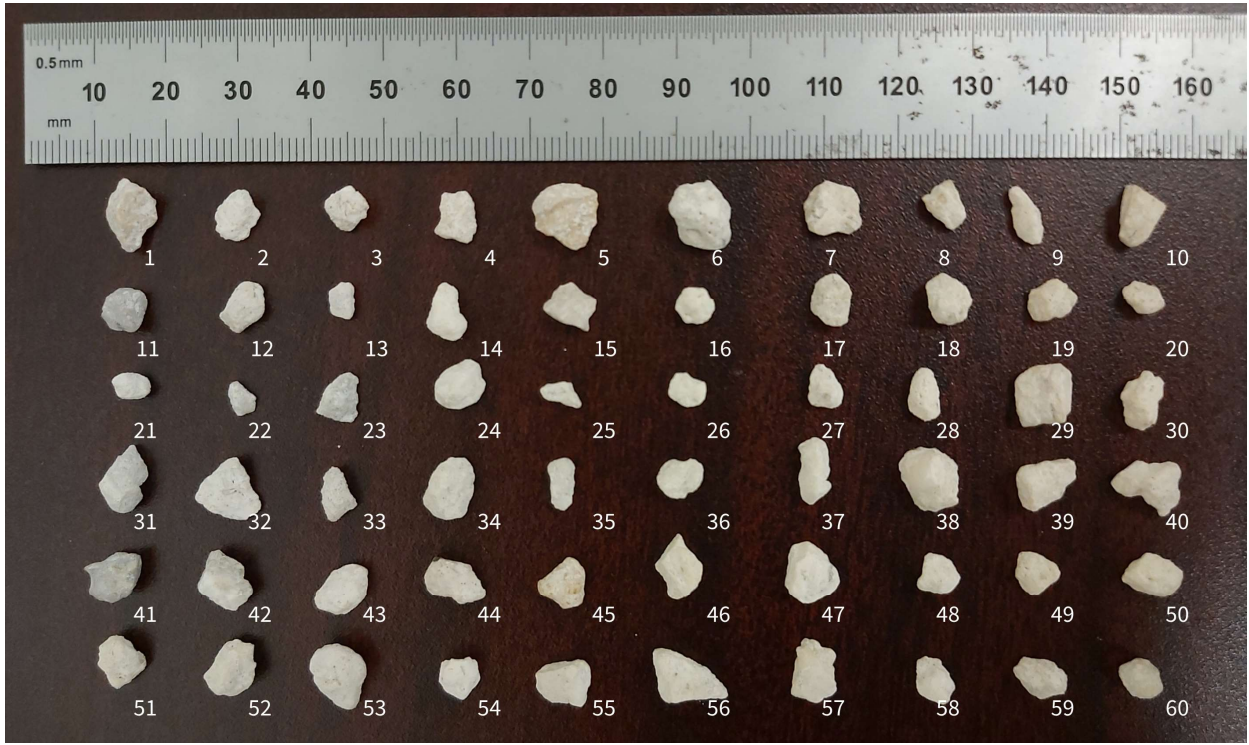

Figure 10. Limestone particles analyzed in this study

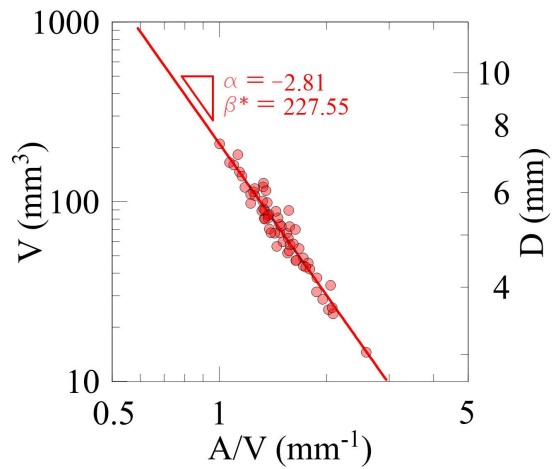

Figure 11. Phenotypic trait of the limestone particles

### 2.3.3  Fragmented Particles

Fragmented particles are analyzed for the evaluation of α and β*. This study analyzes the geometric properties of 86 fragmented particles provided by Zheng et al. [12]. The particle images are shown in Figure 12. The particles with a higher ID are smaller in the figure, and the detailed geometric properties can be found in Table A1 in [12]. Figure 13 shows the phenotypic traits of particle geometries from the power law relation. The obtained α is -2.37 and β* is 271.90. The high β* reflects the high angularity of the fragmented particles due to the sharp edges and corners generated by the breakage. The smaller particles (with a higher ID) are more angular than larger particles, so α = -2.37 accurately represents the shape-size relation.

> Please see Fig. A1 (Morphology of selected grains) in Zheng et al. [12] for Figure 12 (link: https://doi.org/10.1016/j.enggeo.2019.105358). In the journal publication of this manuscript, this figure will be shown with the written permission from Elsevier.

Figure 12. Fragmented particles analyzed in this study

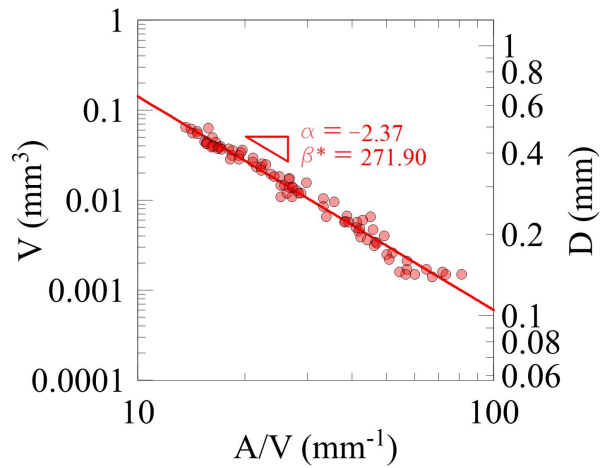

Figure 13. Phenotypic trait of the fragmented particles

## 3   CHARACTERIZATION OF INDIVIDUAL PARTICLE GEOMETRY

Given the A/V and V data of the particles, the phenotypic trait of a granular material (i.e., a group of particles) can be characterized as α and β* as discussed in Section 2. A follow-up question then is whether it is also possible to characterize 'individual' particle shape using A/V and V data. The answer is yes by using Wadell's 'true Sphericity' [13]. Wadell proposed the 'true Sphericity' $S$ to quantify the 3D particle shape in terms of the ratio between two surface areas. Equation (7) is the original definition by Wadell, where A is the surface area of the particle and $A_s$ is the surface area of the reference sphere. The reference sphere is determined so that its volume $V_s$ is equal to the volume V of the particle (i.e., $V_s = V$). The true Sphericity ranges between 0 and 1, with a value close to 1 indicating a near-spherical shape. It is worth noting Wadell also proposed 2D Sphericity and Roundness [6, 13] which are different from the true Sphericity that evaluates the 3D shape.

$$S = A_s / A \qquad (7)$$

The research community has used the original definition presented by Equation (7) nearly for a century without realizing that $S$ can be re-formulated to define the shape as a function of surface area A, volume V, and size D. To be more specific, the inverse of true Sphericity $S^{-1}$ is equal to A/V × D/6 as shown in Equation (8) where D is the diameter of reference sphere having the same volume with the particle. Therefore, the individual particle shape can be estimated in terms of the true Sphericity for any A/V and V data as D can be computed from V using Equation (5).

The derivation of the relation is straightforward: $A_s$ can be expressed in terms of V and D as in Equation (9) since $V = V_s$. Then, combining Equation (7) and (9) gives Equation (8). Figure 14 shows examples of five different shapes with their corresponding A, V, and D values. Since $S^{-1}$ is the inverse of $S$, the lowest possible value is 1 for the sphere (Figure 14a) which increases with angularity: The cube (Figure 14b) has the higher $S^{-1}$ value due to the higher angularity, and the great stellated dodecahedron (Figure 14e) has the highest $S^{-1}$ value due to the very high angularity.

$$1/S = S^{-1} = A/V \times D/6 \qquad (8)$$

$$A_s = 4 \times \pi \times (D/2)^2 = V_s \times 6/D = V \times 6/D \qquad (9)$$

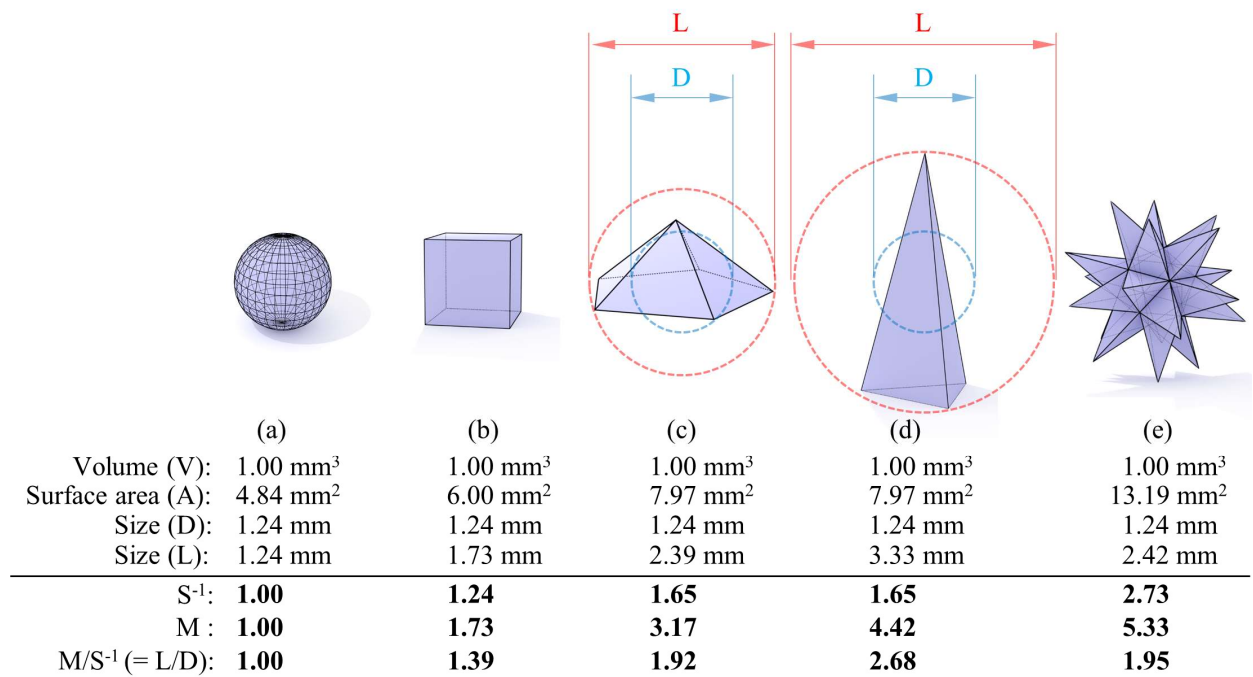

|  | (a) | (b) | (c) | (d) | (e) |
|---|---|---|---|---|---|
| Volume (V): | 1.00 mm³ | 1.00 mm³ | 1.00 mm³ | 1.00 mm³ | 1.00 mm³ |
| Surface area (A): | 4.84 mm² | 6.00 mm² | 7.97 mm² | 7.97 mm² | 13.19 mm² |
| Size (D): | 1.24 mm | 1.24 mm | 1.24 mm | 1.24 mm | 1.24 mm |
| Size (L): | 1.24 mm | 1.73 mm | 2.39 mm | 3.33 mm | 2.42 mm |
| $S^{-1}$: | 1.00 | 1.24 | 1.65 | 1.65 | 2.73 |
| M : | 1.00 | 1.73 | 3.17 | 4.42 | 5.33 |
| $M/S^{-1}$ (= L/D): | 1.00 | 1.39 | 1.92 | 2.68 | 1.95 |

Figure 14. Particle shape characterization in terms of surface area, volume, and size: (a) Sphere; (b) Cube; (c) Pentagon pyramid; (d) Elongated tetrahedron; and (e) Great stellated dodecahedron

However, there is a limitation to the use of true Sphericity as a sole 3D particle shape index, because it is influenced by both particle elongation and angularity [14, 15]. The particle elongation is concerned about the overall form that is defined at the particle's diameter scale O(*d*), e.g., sphere vs. ellipsoid. On the other hand, the angularity is related to the corner sharpness defined at a smaller length scale by an order of magnitude O(*d*/10), e.g., sphere vs. cube [6, 16]. Since the true Sphericity is affected by both elongation and angularity, different shapes may have a same $S^{-1}$ value such as Figure 14c and d: the pentagon pyramid (Figure 14c) has overall high corner angularity whereas the tetrahedron is more elongated (Figure 14d). For this reason, at least additional information regarding elongation is needed to better characterize the shape. The true Sphericity has an inherent limitation in properly evaluating the elongation because the size D in Equation (8) is not the actual particle size (except sphere) but the size of the reference sphere of same volume as the particle. Therefore, all particles in Figure 14 have the same size D regardless of the actual elongation because all particles have the same volume. Since D can be computed from V per Equation (5), the D is also redundant information, thus can be replaced with more meaningful size information to account for the elongation.

Therefore, we propose to use *M* as the second index that is defined with the additional size information L as in Equation (10). The shape index *M* was originally proposed by Su et al. [1] to quantify the particle shape as a function of surface area, volume, and size, then without knowing that Wadell's true sphericity could be similarly formulated. Compared to Equation (8), the shape index *M* considers L instead of D. The size L is the actual particle size evaluated by the circumdiameter (i.e., diameter of particle's circumsphere) which therefore considers the length to the farthest corner of the particle. A schematic comparison of D and L is also shown in Figure 14c and d, where D represents the diameter of sphere in Figure 14a. Since L is always greater than or equal to D, the *M* values obtained for the five shapes in Figure 14 are greater than or equal to $S^{-1}$. Therefore, as in Equation (11), the $M/S^{-1}$ ratio (simply $M \times S$) is the ratio of L/D which can distinctively inform the particle elongation.

$$M = A/V \times L/6 \tag{10}$$

$$M/S^{-1} (= M \times S) = L/D \tag{11}$$

The elongated tetrahedron (Figure 14d) has the higher $M/S^{-1}$ value than the pentagon pyramid (Figure 14c) because of the elongation. The great stellated dodecahedron (Figure 14e) has the highest values of *M* and $S^{-1}$ among the five shapes due to the very high angularity, but the $M/S^{-1}$ value is lower than that of the elongated tetrahedron. In fact, if we can compute *M* using Equation (10), it means we can also get L/D because D can be obtained from V using Equation (5). Therefore, *M* also contains the $S^{-1}$ information. The $S^{-1}$ and $M$ values of the example particles in Section 2.3 are summarized in Table A.6 and Table A.7 for interested readers.

The A/V and V data can be presented as a 3D plot with the additional information L, on which a power regression can be also made. For example, Figure 15 shows the phenotypic traits of the 200 polyhedral particles (in Section 2.3.1) as a 3D plot.

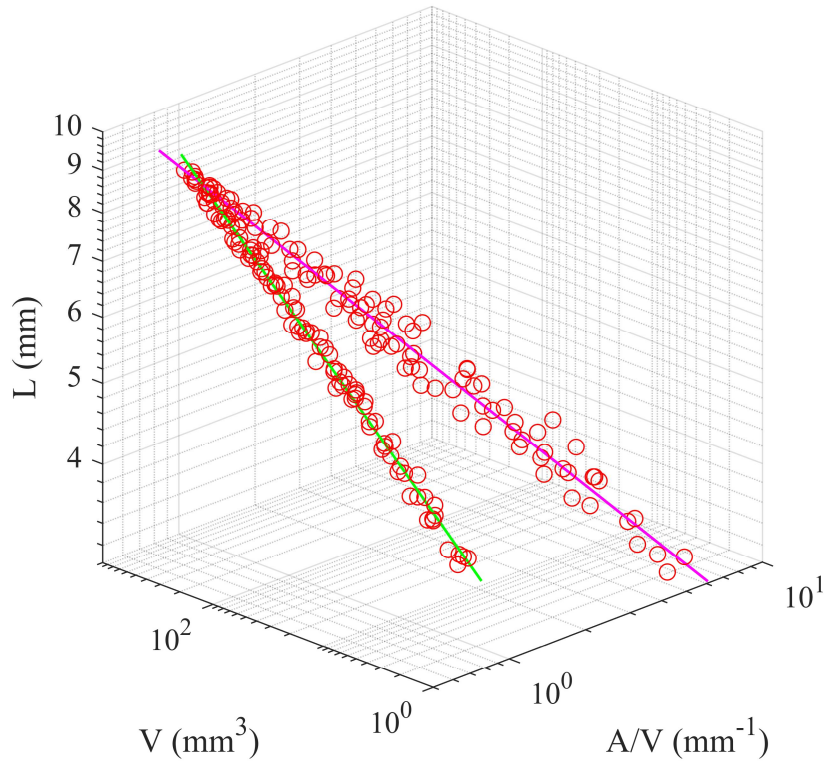

Figure 15. Two different phenotypic traits in the 200 particles evaluated in 3D

## 4  CONCLUDING REMARKS

This paper provides a new perspective on how the phenotypic trait can be discovered in the particle geometries, demonstrating the particle surface-area-to-volume ratio (A/V) and the particle volume (V) are the key information. The proposed approach using A/V and V provides a unified method that can comprehensively characterize the particle geometry at multiple scales from (i) granular material to (ii) single particle.

(i) Phenotypic trait of granular material - The relation between A/V and V data of a granular material can be approximated by a 'power-law' that can uncover the 'phenotypic trait' of the particle geometries.

This phenotypic trait is realized as a linear plot in a log-log space, where the power value α (slope) represents the relation between particle shape and size (i.e., variation), and the intercept term β* (evaluated by the constrained analysis for α = -3) informs the angularity of the average shape. In addition, particle volume, surface area, and size can be considered together in this space, allowing for comprehensive particle geometry characterization.

(ii) Single particle geometry - The A/V and V information can be also used for single particle geometry quantification. The 3D shape of individual particle can be characterized using the A/V and V values in terms of Wadell's true Sphericity *S*. The concept is linked to another shape index *M* that extends the concept of Wadell's true Sphericity with additional size information L. This paper finds $M/S^{-1}$ provides the useful information about particle elongation. Furthermore, this approach comprehensively describes how the particle shape is related to surface area (A), volume (V), and size (D or L) which is a powerful feature compared to the conventional methods (e.g., 2D sphericity and roundness) that separately describe the shape from the surface area, volume, and size.

The findings in this study can be extended to identify the phenotypic trait of any granular materials in general, e.g., coffee beans, cereal grains, chemical powders, etc. The granular materials affect our daily lives in every different way considering it is the second-most manipulated material next to water [17]. This study will help to systematically address the particle geometry and better understand the complex behavior of the granular materials.

## ACKNOWLEDGEMENTS

This work is sponsored in part by the US National Science Foundation under the awards CMMI #1938431 and #1938285. The opinions, findings, conclusions, or recommendations expressed in this article are solely those of the authors and do not necessarily reflect the views of the funding agency. The authors would like to thank Mr. Michael Kim of the Florida Department of Transportation District Four and Six Materials Office for providing the Florida limestone construction aggregate used in this study.

# COMPLIANCE WITH ETHICAL STANDARDS

Conflicts of interest

The authors declare that they have no conflict of interest.

Statement of human and animal rights

This research did not involve human participants or animals.

# APPENDIX

Table A.1. Geometric properties of Group 1; The power regression yields α = -3 and β = 113.09 (= β*)

| Group 1 | D (mm) | V (mm$^3$) | A (mm$^2$) | A/V (mm$^{-1}$) | log(V) | log(A/V) |
|---|---|---|---|---|---|---|
| | 0.322 | 0.018 | 0.327 | 18.611 | -1.756 | 1.270 |
| | 0.403 | 0.034 | 0.510 | 14.888 | -1.465 | 1.173 |
| | 0.484 | 0.059 | 0.735 | 12.407 | -1.228 | 1.094 |
| | 0.564 | 0.094 | 1.000 | 10.635 | -1.027 | 1.027 |
| | 0.645 | 0.140 | 1.306 | 9.305 | -0.853 | 0.969 |
| | 0.725 | 0.200 | 1.653 | 8.271 | -0.699 | 0.918 |
| | 0.806 | 0.274 | 2.041 | 7.444 | -0.562 | 0.872 |
| Sphere | 1.612 | 2.193 | 8.163 | 3.722 | 0.341 | 0.571 |
| | 2.418 | 7.402 | 18.368 | 2.481 | 0.869 | 0.395 |
| | 3.224 | 17.546 | 32.654 | 1.861 | 1.244 | 0.270 |
| | 4.030 | 34.269 | 51.022 | 1.489 | 1.535 | 0.173 |
| | 4.836 | 59.218 | 73.471 | 1.241 | 1.772 | 0.094 |
| | 5.642 | 94.035 | 100.003 | 1.063 | 1.973 | 0.027 |
| | 6.448 | 140.368 | 130.616 | 0.931 | 2.147 | -0.031 |
| | 7.254 | 199.859 | 165.311 | 0.827 | 2.301 | -0.082 |
| Average | | | | | 0.306 | 0.582 |

Table A.2. Geometric properties of Group 2; The power regression yields α = -3 and β = 216 (= β*)

| Group 2 | D (mm) | V (mm$^3$) | A (mm$^2$) | A/V (mm$^{-1}$) | log(V) | log(A/V) |
|---|---|---|---|---|---|---|
| | 0.400 | 0.034 | 0.624 | 18.611 | -1.475 | 1.270 |
| | 0.500 | 0.065 | 0.974 | 14.888 | -1.184 | 1.173 |
| | 0.600 | 0.113 | 1.403 | 12.407 | -0.947 | 1.094 |
| | 0.700 | 0.180 | 1.910 | 10.635 | -0.746 | 1.027 |
| | 0.800 | 0.268 | 2.495 | 9.305 | -0.572 | 0.969 |
| | 0.900 | 0.382 | 3.157 | 8.271 | -0.418 | 0.918 |
| | 1.000 | 0.524 | 3.898 | 7.444 | -0.281 | 0.872 |
| Cube | 2.000 | 4.189 | 15.591 | 3.722 | 0.622 | 0.571 |
| | 3.000 | 14.137 | 35.080 | 2.481 | 1.150 | 0.395 |
| | 4.000 | 33.510 | 62.364 | 1.861 | 1.525 | 0.270 |
| | 5.000 | 65.450 | 97.444 | 1.489 | 1.816 | 0.173 |
| | 6.000 | 113.097 | 140.320 | 1.241 | 2.053 | 0.094 |
| | 7.000 | 179.594 | 190.991 | 1.063 | 2.254 | 0.027 |
| | 8.000 | 268.083 | 249.458 | 0.931 | 2.428 | -0.031 |
| | 9.000 | 381.704 | 315.720 | 0.827 | 2.582 | -0.082 |
| Average | | | | | 0.587 | 0.582 |

Table A.3. Geometric properties of Group 3; The power regression yields α = -3 and β = 374.12 (= β*)

| Group 3 | D (mm) | V (mm³) | A (mm²) | A/V (mm⁻¹) | log(V) | log(A/V) |
|---|---|---|---|---|---|---|
| | 0.480 | 0.058 | 1.080 | 18.611 | -1.236 | 1.270 |
| | 0.600 | 0.113 | 1.688 | 14.888 | -0.946 | 1.173 |
| | 0.721 | 0.196 | 2.430 | 12.407 | -0.708 | 1.094 |
| | 0.841 | 0.311 | 3.308 | 10.635 | -0.507 | 1.027 |
| | 0.961 | 0.464 | 4.321 | 9.305 | -0.333 | 0.969 |
| | 1.081 | 0.661 | 5.468 | 8.271 | -0.180 | 0.918 |
| | 1.201 | 0.907 | 6.751 | 7.444 | -0.042 | 0.872 |
| Tetrahedron | 2.402 | 7.255 | 27.005 | 3.722 | 0.861 | 0.571 |
| | 3.603 | 24.486 | 60.760 | 2.481 | 1.389 | 0.395 |
| | 4.804 | 58.042 | 108.018 | 1.861 | 1.764 | 0.270 |
| | 6.005 | 113.362 | 168.779 | 1.489 | 2.054 | 0.173 |
| | 7.206 | 195.890 | 243.041 | 1.241 | 2.292 | 0.094 |
| | 8.407 | 311.067 | 330.806 | 1.063 | 2.493 | 0.027 |
| | 9.607 | 464.333 | 432.073 | 0.931 | 2.667 | -0.031 |
| | 10.808 | 661.130 | 546.843 | 0.827 | 2.820 | -0.082 |
| Average | | | | | 0.826 | 0.582 |

Table A.4 Geometric properties of Group A and B in Figure 5

| Group A | D (mm) | V (mm³) | A (mm²) | A/V (mm⁻¹) | log(V) | log(A/V) | |
|---|---|---|---|---|---|---|---|
| Sphere | 2.418 | 7.402 | 18.368 | 2.481 | 0.869 | 0.395 | |
| Cube | 4.000 | 33.510 | 62.364 | 1.861 | 1.525 | 0.270 | |
| Tetrahedron | 6.005 | 113.362 | 168.779 | 1.489 | 2.054 | 0.173 | |
| Average | | | | | 1.483 | 0.279 | β* = 209.08 |
| Group B | D (mm) | V (mm³) | A (mm²) | A/V (mm⁻¹) | log(V) | log(A/V) | |
| Sphere | 4.836 | 59.218 | 73.471 | 1.241 | 1.772 | 0.094 | |
| Cube | 4.000 | 33.510 | 62.364 | 1.861 | 1.525 | 0.270 | |
| Tetrahedron | 3.603 | 24.486 | 60.760 | 2.481 | 1.389 | 0.395 | |
| Average | | | | | 1.562 | 0.253 | β* = 209.08 |
| Group B | D (cm) | V (cm³) | A (cm²) | A/V (cm⁻¹) | log(V) | log(A/V) | |
| Sphere | 0.484 | 0.059 | 0.735 | 12.407 | -1.228 | 1.094 | |
| Cube | 0.400 | 0.034 | 0.624 | 18.611 | -1.475 | 1.270 | |
| Tetrahedron | 0.360 | 0.024 | 0.608 | 24.814 | -1.611 | 1.395 | |
| Average | | | | | -1.438 | 1.253 | β* = 209.08 |

Table A.5. Geometric properties of particles in Figure 6; All particles have a unit volume

| Figure 6a | D (mm) | V (mm³) | A (mm²) | A/V (mm⁻¹) | log(V) | log(A/V) | |
|---|---|---|---|---|---|---|---|
| Sphere | 1.241 | 1.000 | 4.836 | 4.836 | 0.000 | 0.684 | |
| Cube | 1.241 | 1.000 | 6.000 | 6.000 | 0.000 | 0.778 | |
| Tetrahedron | 1.241 | 1.000 | 7.206 | 7.206 | 0.000 | 0.858 | |
| Average | | | | | 0.000 | 0.773 | β* = 209.08 |
| Figure 6b | D (mm) | V (mm³) | A (mm²) | A/V (mm⁻¹) | log(V) | log(A/V) | |
| Sphere | 1.241 | 1.000 | 4.836 | 4.836 | 0.000 | 0.684 | |
| Cube | 1.241 | 1.000 | 6.000 | 6.000 | 0.000 | 0.778 | |
| ET* | 1.241 | 1.000 | 7.972 | 7.972 | 0.000 | 0.902 | |
| Average | | | | | 0.000 | 0.788 | β* = 231.31 |
| Figure 6c | D (mm) | V (mm³) | A (mm²) | A/V (mm⁻¹) | log(V) | log(A/V) | |
| Sphere 1 | 1.241 | 1.000 | 4.836 | 4.836 | 0.000 | 0.684 | |
| Sphere 2 | 1.241 | 1.000 | 4.836 | 4.836 | 0.000 | 0.684 | |
| Cube | 1.241 | 1.000 | 6.000 | 6.000 | 0.000 | 0.778 | |
| Tetrahedron | 1.241 | 1.000 | 7.206 | 7.206 | 0.000 | 0.858 | |
| Average | | | | | 0.000 | 0.751 | β* = 179.31 |
| Figure 6d | D (mm) | V (mm³) | A (mm²) | A/V (mm⁻¹) | log(V) | log(A/V) | |
| Sphere 1 | 1.241 | 1.000 | 4.836 | 4.836 | 0.000 | 0.684 | |
| Sphere 2 | 1.241 | 1.000 | 4.836 | 4.836 | 0.000 | 0.684 | |
| … | … | … | … | … | … | … | |
| Sphere 100 | 1.241 | 1.000 | 4.836 | 4.836 | 0.000 | 0.684 | |
| Cube | 1.241 | 1.000 | 6.000 | 6.000 | 0.000 | 0.778 | |
| Tetrahedron | 1.241 | 1.000 | 7.206 | 7.206 | 0.000 | 0.858 | |
| Average | | | | | 0.000 | 0.687 | β* = 115.16 |

*ET: Elongated Tetrahedron

Table A.6. Geometric properties of 200 polyhedral particles

| ID | D (mm) | V (mm³) | A (mm²) | A/V | log(V) | log(A/V) | L (mm) | $S^{-1}$ | M |
|---|---|---|---|---|---|---|---|---|---|
| 1 | 8.550 | 327.290 | 232.718 | 0.711 | 2.515 | -0.148 | 8.984 | 1.013 | 1.065 |
| 2 | 8.487 | 320.079 | 228.506 | 0.714 | 2.505 | -0.146 | 8.876 | 1.010 | 1.056 |
| 3 | 8.510 | 322.706 | 228.806 | 0.709 | 2.509 | -0.149 | 8.830 | 1.006 | 1.043 |
| 4 | 8.069 | 275.090 | 207.197 | 0.753 | 2.439 | -0.123 | 8.780 | 1.013 | 1.102 |
| 5 | 8.166 | 285.101 | 211.747 | 0.743 | 2.455 | -0.129 | 8.726 | 1.011 | 1.080 |
| 6 | 8.148 | 283.212 | 210.673 | 0.744 | 2.452 | -0.129 | 8.612 | 1.010 | 1.068 |
| 7 | 8.035 | 271.628 | 204.718 | 0.754 | 2.434 | -0.123 | 8.604 | 1.009 | 1.081 |
| 8 | 8.037 | 271.849 | 204.696 | 0.753 | 2.434 | -0.123 | 8.565 | 1.009 | 1.075 |
| 9 | 8.198 | 288.523 | 212.411 | 0.736 | 2.460 | -0.133 | 8.501 | 1.006 | 1.043 |
| 10 | 7.922 | 260.339 | 199.146 | 0.765 | 2.416 | -0.116 | 8.435 | 1.010 | 1.075 |
| 11 | 8.115 | 279.769 | 208.054 | 0.744 | 2.447 | -0.129 | 8.370 | 1.006 | 1.037 |
| 12 | 8.152 | 283.699 | 211.156 | 0.744 | 2.453 | -0.128 | 8.365 | 1.011 | 1.038 |

| 13 | 7.512 | 221.988 | 178.867 | 0.806 | 2.346 | -0.094 | 8.262 | 1.009 | 1.109 |
|---|---|---|---|---|---|---|---|---|---|
| 14 | 7.883 | 256.509 | 196.280 | 0.765 | 2.409 | -0.116 | 8.130 | 1.005 | 1.037 |
| 15 | 7.751 | 243.775 | 189.881 | 0.779 | 2.387 | -0.109 | 8.020 | 1.006 | 1.041 |
| 16 | 7.596 | 229.494 | 182.857 | 0.797 | 2.361 | -0.099 | 8.020 | 1.009 | 1.065 |
| 17 | 7.109 | 188.155 | 161.096 | 0.856 | 2.275 | -0.067 | 8.010 | 1.015 | 1.143 |
| 18 | 7.495 | 220.458 | 178.369 | 0.809 | 2.343 | -0.092 | 7.996 | 1.011 | 1.078 |
| 19 | 7.682 | 237.395 | 186.920 | 0.787 | 2.375 | -0.104 | 7.979 | 1.008 | 1.047 |
| 20 | 7.397 | 211.951 | 173.829 | 0.820 | 2.326 | -0.086 | 7.870 | 1.011 | 1.076 |
| 21 | 7.209 | 196.139 | 165.054 | 0.842 | 2.293 | -0.075 | 7.843 | 1.011 | 1.100 |
| 22 | 7.147 | 191.186 | 161.606 | 0.845 | 2.281 | -0.073 | 7.635 | 1.007 | 1.076 |
| 23 | 7.349 | 207.783 | 171.059 | 0.823 | 2.318 | -0.084 | 7.601 | 1.008 | 1.043 |
| 24 | 7.278 | 201.886 | 167.276 | 0.829 | 2.305 | -0.082 | 7.539 | 1.005 | 1.041 |
| 25 | 7.173 | 193.269 | 162.905 | 0.843 | 2.286 | -0.074 | 7.499 | 1.008 | 1.053 |
| 26 | 6.793 | 164.157 | 147.046 | 0.896 | 2.215 | -0.048 | 7.436 | 1.014 | 1.110 |
| 27 | 6.865 | 169.398 | 150.063 | 0.886 | 2.229 | -0.053 | 7.383 | 1.014 | 1.090 |
| 28 | 7.135 | 190.218 | 160.938 | 0.846 | 2.279 | -0.073 | 7.384 | 1.006 | 1.041 |
| 29 | 6.594 | 150.141 | 138.827 | 0.925 | 2.177 | -0.034 | 7.379 | 1.016 | 1.137 |
| 30 | 6.841 | 167.654 | 148.309 | 0.885 | 2.224 | -0.053 | 7.291 | 1.009 | 1.075 |
| 31 | 6.793 | 164.120 | 146.524 | 0.893 | 2.215 | -0.049 | 7.280 | 1.011 | 1.083 |
| 32 | 6.603 | 150.721 | 138.425 | 0.918 | 2.178 | -0.037 | 7.267 | 1.011 | 1.112 |
| 33 | 6.891 | 171.323 | 149.775 | 0.874 | 2.234 | -0.058 | 7.224 | 1.004 | 1.053 |
| 34 | 6.686 | 156.489 | 141.858 | 0.907 | 2.194 | -0.043 | 7.161 | 1.010 | 1.082 |
| 35 | 6.549 | 147.044 | 135.700 | 0.923 | 2.167 | -0.035 | 7.033 | 1.007 | 1.082 |
| 36 | 6.566 | 148.238 | 136.005 | 0.917 | 2.171 | -0.037 | 6.976 | 1.004 | 1.067 |
| 37 | 6.476 | 142.207 | 133.139 | 0.936 | 2.153 | -0.029 | 6.967 | 1.011 | 1.087 |
| 38 | 6.419 | 138.474 | 130.285 | 0.941 | 2.141 | -0.026 | 6.871 | 1.007 | 1.077 |
| 39 | 6.219 | 125.946 | 122.711 | 0.974 | 2.100 | -0.011 | 6.772 | 1.010 | 1.100 |
| 40 | 5.995 | 112.798 | 113.789 | 1.009 | 2.052 | 0.004 | 6.740 | 1.008 | 1.133 |
| 41 | 6.223 | 126.178 | 122.993 | 0.975 | 2.101 | -0.011 | 6.741 | 1.011 | 1.095 |
| 42 | 6.309 | 131.495 | 126.011 | 0.958 | 2.119 | -0.019 | 6.718 | 1.008 | 1.073 |
| 43 | 6.177 | 123.382 | 121.169 | 0.982 | 2.091 | -0.008 | 6.717 | 1.011 | 1.100 |
| 44 | 5.815 | 102.934 | 107.128 | 1.041 | 2.013 | 0.017 | 6.544 | 1.009 | 1.135 |
| 45 | 6.067 | 116.915 | 116.542 | 0.997 | 2.068 | -0.001 | 6.527 | 1.008 | 1.084 |
| 46 | 5.781 | 101.145 | 105.924 | 1.047 | 2.005 | 0.020 | 6.345 | 1.009 | 1.107 |
| 47 | 5.711 | 97.509 | 103.161 | 1.058 | 1.989 | 0.024 | 6.314 | 1.007 | 1.113 |
| 48 | 5.978 | 111.839 | 113.060 | 1.011 | 2.049 | 0.005 | 6.296 | 1.007 | 1.061 |
| 49 | 5.816 | 102.983 | 106.821 | 1.037 | 2.013 | 0.016 | 6.050 | 1.005 | 1.046 |
| 50 | 5.681 | 96.024 | 102.063 | 1.063 | 1.982 | 0.026 | 6.026 | 1.006 | 1.067 |
| 51 | 5.573 | 90.624 | 98.316 | 1.085 | 1.957 | 0.035 | 6.007 | 1.008 | 1.086 |
| 52 | 5.485 | 86.386 | 95.043 | 1.100 | 1.936 | 0.041 | 5.954 | 1.006 | 1.092 |
| 53 | 5.640 | 93.912 | 100.374 | 1.069 | 1.973 | 0.029 | 5.935 | 1.005 | 1.057 |
| 54 | 5.404 | 82.631 | 93.016 | 1.126 | 1.917 | 0.051 | 5.931 | 1.014 | 1.113 |
| 55 | 5.504 | 87.290 | 96.052 | 1.100 | 1.941 | 0.042 | 5.915 | 1.009 | 1.085 |
| 56 | 5.217 | 74.343 | 86.582 | 1.165 | 1.871 | 0.066 | 5.861 | 1.013 | 1.138 |
| 57 | 5.209 | 73.993 | 86.166 | 1.165 | 1.869 | 0.066 | 5.708 | 1.011 | 1.108 |
| 58 | 5.103 | 69.564 | 82.545 | 1.187 | 1.842 | 0.074 | 5.695 | 1.009 | 1.126 |
| 59 | 5.022 | 66.328 | 80.169 | 1.209 | 1.822 | 0.082 | 5.592 | 1.012 | 1.127 |
| 60 | 5.287 | 77.360 | 88.307 | 1.142 | 1.889 | 0.057 | 5.484 | 1.006 | 1.043 |
| 61 | 4.927 | 62.641 | 77.038 | 1.230 | 1.797 | 0.090 | 5.386 | 1.010 | 1.104 |
| 62 | 4.943 | 63.249 | 77.885 | 1.231 | 1.801 | 0.090 | 5.343 | 1.015 | 1.097 |
| 63 | 4.847 | 59.640 | 74.805 | 1.254 | 1.776 | 0.098 | 5.345 | 1.013 | 1.117 |
| 64 | 4.749 | 56.097 | 71.986 | 1.283 | 1.749 | 0.108 | 5.265 | 1.016 | 1.126 |
| 65 | 4.818 | 58.545 | 73.572 | 1.257 | 1.767 | 0.099 | 5.199 | 1.009 | 1.089 |
| 66 | 4.643 | 52.423 | 68.262 | 1.302 | 1.720 | 0.115 | 5.170 | 1.008 | 1.122 |
| 67 | 4.774 | 56.974 | 72.619 | 1.275 | 1.756 | 0.105 | 5.146 | 1.014 | 1.093 |
| 68 | 4.882 | 60.920 | 75.246 | 1.235 | 1.785 | 0.092 | 5.113 | 1.005 | 1.053 |
| 69 | 4.533 | 48.764 | 65.490 | 1.343 | 1.688 | 0.128 | 5.095 | 1.015 | 1.140 |
| 70 | 4.510 | 48.031 | 64.445 | 1.342 | 1.682 | 0.128 | 5.058 | 1.009 | 1.131 |
| 71 | 4.529 | 48.640 | 64.995 | 1.336 | 1.687 | 0.126 | 5.034 | 1.009 | 1.121 |
| 72 | 4.531 | 48.700 | 65.033 | 1.335 | 1.688 | 0.126 | 4.983 | 1.008 | 1.109 |
| 73 | 4.595 | 50.783 | 66.997 | 1.319 | 1.706 | 0.120 | 4.967 | 1.010 | 1.092 |
| 74 | 4.373 | 43.794 | 60.958 | 1.392 | 1.641 | 0.144 | 4.939 | 1.015 | 1.146 |
| 75 | 4.380 | 44.003 | 60.977 | 1.386 | 1.643 | 0.142 | 4.878 | 1.012 | 1.127 |
| 76 | 4.288 | 41.272 | 58.230 | 1.411 | 1.616 | 0.149 | 4.682 | 1.008 | 1.101 |
| 77 | 4.191 | 38.539 | 55.780 | 1.447 | 1.586 | 0.161 | 4.687 | 1.011 | 1.131 |
| 78 | 4.278 | 40.989 | 57.920 | 1.413 | 1.613 | 0.150 | 4.617 | 1.007 | 1.087 |
| 79 | 3.911 | 31.321 | 48.455 | 1.547 | 1.496 | 0.189 | 4.445 | 1.008 | 1.146 |
| 80 | 4.044 | 34.637 | 51.951 | 1.500 | 1.540 | 0.176 | 4.415 | 1.011 | 1.104 |
| 81 | 4.023 | 34.103 | 51.101 | 1.498 | 1.533 | 0.176 | 4.382 | 1.005 | 1.094 |
| 82 | 4.072 | 35.363 | 52.397 | 1.482 | 1.549 | 0.171 | 4.343 | 1.006 | 1.072 |
| 83 | 3.944 | 32.127 | 49.420 | 1.538 | 1.507 | 0.187 | 4.266 | 1.011 | 1.094 |
| 84 | 3.791 | 28.530 | 45.526 | 1.596 | 1.455 | 0.203 | 4.153 | 1.008 | 1.105 |
| 85 | 3.829 | 29.401 | 46.518 | 1.582 | 1.468 | 0.199 | 4.090 | 1.010 | 1.079 |
| 86 | 3.735 | 27.282 | 44.314 | 1.624 | 1.436 | 0.211 | 4.074 | 1.011 | 1.103 |
| 87 | 3.549 | 23.414 | 40.119 | 1.713 | 1.369 | 0.234 | 4.061 | 1.014 | 1.160 |
| 88 | 3.646 | 25.368 | 42.016 | 1.656 | 1.404 | 0.219 | 3.832 | 1.006 | 1.058 |
| 89 | 3.545 | 23.323 | 39.891 | 1.710 | 1.368 | 0.233 | 3.826 | 1.010 | 1.091 |
| 90 | 3.448 | 21.462 | 37.783 | 1.760 | 1.332 | 0.246 | 3.824 | 1.012 | 1.122 |
| 91 | 3.313 | 19.034 | 34.795 | 1.828 | 1.280 | 0.262 | 3.749 | 1.009 | 1.142 |
| 92 | 3.309 | 18.976 | 34.743 | 1.831 | 1.278 | 0.263 | 3.646 | 1.010 | 1.113 |
| 93 | 3.335 | 19.418 | 35.187 | 1.812 | 1.288 | 0.258 | 3.610 | 1.007 | 1.089 |
| 94 | 3.334 | 19.396 | 35.280 | 1.819 | 1.288 | 0.260 | 3.592 | 1.011 | 1.090 |
| 95 | 3.407 | 20.709 | 36.784 | 1.776 | 1.316 | 0.249 | 3.597 | 1.009 | 1.065 |
| 96 | 3.137 | 16.163 | 31.091 | 1.924 | 1.209 | 0.284 | 3.323 | 1.006 | 1.065 |
| 97 | 3.006 | 14.222 | 28.597 | 2.011 | 1.153 | 0.303 | 3.285 | 1.007 | 1.101 |
| 98 | 2.962 | 13.613 | 27.910 | 2.050 | 1.134 | 0.312 | 3.265 | 1.012 | 1.116 |
| 99 | 2.905 | 12.838 | 26.798 | 2.087 | 1.109 | 0.320 | 3.252 | 1.011 | 1.131 |
| 100 | 3.019 | 14.413 | 28.805 | 1.999 | 1.159 | 0.301 | 3.192 | 1.006 | 1.063 |
| 101 | 7.437 | 215.333 | 181.430 | 0.843 | 2.333 | -0.074 | 8.109 | 1.044 | 1.139 |
| 102 | 8.124 | 280.709 | 214.140 | 0.763 | 2.448 | -0.118 | 8.719 | 1.033 | 1.109 |
| 103 | 7.593 | 229.203 | 185.476 | 0.809 | 2.360 | -0.092 | 8.042 | 1.024 | 1.085 |
| 104 | 8.547 | 326.946 | 232.899 | 0.712 | 2.514 | -0.147 | 8.912 | 1.015 | 1.058 |
| 105 | 7.707 | 239.696 | 191.451 | 0.799 | 2.380 | -0.098 | 8.296 | 1.026 | 1.104 |
| 106 | 5.992 | 112.634 | 122.463 | 1.087 | 2.052 | 0.036 | 6.954 | 1.086 | 1.260 |
| 107 | 7.402 | 212.349 | 179.450 | 0.845 | 2.327 | -0.073 | 8.167 | 1.043 | 1.150 |
| 108 | 8.094 | 277.643 | 209.486 | 0.755 | 2.443 | -0.122 | 8.587 | 1.018 | 1.080 |
| 109 | 8.887 | 367.542 | 250.045 | 0.680 | 2.565 | -0.167 | 9.135 | 1.008 | 1.036 |
| 110 | 8.613 | 334.526 | 234.685 | 0.702 | 2.524 | -0.154 | 9.101 | 1.007 | 1.064 |
| 111 | 4.494 | 47.538 | 78.938 | 1.661 | 1.677 | 0.220 | 5.920 | 1.244 | 1.639 |
| 112 | 3.844 | 29.752 | 54.514 | 1.832 | 1.474 | 0.263 | 5.566 | 1.174 | 1.700 |
| 113 | 5.608 | 92.344 | 112.758 | 1.221 | 1.965 | 0.087 | 6.850 | 1.141 | 1.394 |
| 114 | 4.429 | 45.501 | 67.742 | 1.489 | 1.658 | 0.173 | 5.822 | 1.099 | 1.445 |
| 115 | 5.622 | 93.016 | 105.239 | 1.131 | 1.969 | 0.054 | 6.915 | 1.060 | 1.304 |
| 116 | 6.865 | 169.370 | 154.096 | 0.910 | 2.229 | -0.041 | 8.168 | 1.041 | 1.239 |
| 117 | 7.127 | 189.581 | 164.399 | 0.867 | 2.278 | -0.062 | 8.201 | 1.030 | 1.185 |
| 118 | 7.429 | 214.690 | 176.627 | 0.823 | 2.332 | -0.085 | 8.264 | 1.019 | 1.133 |
| 119 | 7.701 | 239.175 | 188.767 | 0.789 | 2.379 | -0.103 | 8.395 | 1.013 | 1.104 |
| 120 | 7.819 | 250.261 | 194.500 | 0.777 | 2.398 | -0.109 | 8.623 | 1.013 | 1.117 |
| 121 | 4.619 | 51.605 | 73.095 | 1.416 | 1.713 | 0.151 | 6.178 | 1.090 | 1.458 |
| 122 | 5.815 | 102.953 | 112.975 | 1.097 | 2.013 | 0.040 | 7.395 | 1.063 | 1.352 |
| 123 | 4.414 | 45.042 | 68.136 | 1.513 | 1.654 | 0.180 | 6.049 | 1.113 | 1.525 |
| 124 | 3.705 | 26.632 | 49.764 | 1.869 | 1.425 | 0.272 | 5.119 | 1.154 | 1.594 |
| 125 | 6.525 | 145.441 | 137.900 | 0.948 | 2.163 | -0.023 | 7.563 | 1.031 | 1.195 |
| 126 | 5.518 | 87.994 | 98.683 | 1.121 | 1.944 | 0.050 | 6.853 | 1.031 | 1.281 |
| 127 | 6.333 | 133.011 | 128.502 | 0.966 | 2.124 | -0.015 | 7.496 | 1.020 | 1.207 |
| 128 | 8.651 | 338.957 | 237.502 | 0.701 | 2.530 | -0.154 | 8.941 | 1.010 | 1.044 |
| 129 | 6.778 | 163.018 | 148.063 | 0.908 | 2.212 | -0.042 | 7.851 | 1.026 | 1.188 |
| 130 | 7.462 | 217.531 | 179.324 | 0.824 | 2.338 | -0.084 | 8.458 | 1.025 | 1.162 |
| 131 | 3.717 | 26.885 | 49.693 | 1.848 | 1.430 | 0.267 | 5.343 | 1.145 | 1.646 |
| 132 | 3.944 | 32.129 | 60.506 | 1.883 | 1.507 | 0.275 | 5.886 | 1.238 | 1.848 |
| 133 | 3.779 | 28.258 | 49.239 | 1.742 | 1.451 | 0.241 | 5.566 | 1.097 | 1.617 |
| 134 | 6.440 | 139.842 | 136.088 | 0.973 | 2.146 | -0.012 | 7.850 | 1.045 | 1.273 |
| 135 | 5.277 | 76.954 | 95.094 | 1.236 | 1.886 | 0.092 | 6.918 | 1.087 | 1.425 |
| 136 | 5.987 | 112.388 | 119.050 | 1.059 | 2.051 | 0.025 | 7.398 | 1.057 | 1.306 |
| 137 | 5.923 | 108.777 | 116.311 | 1.069 | 2.037 | 0.029 | 7.171 | 1.055 | 1.278 |
| 138 | 6.913 | 172.970 | 155.490 | 0.899 | 2.238 | -0.046 | 8.041 | 1.036 | 1.205 |
| 139 | 7.553 | 225.597 | 182.337 | 0.808 | 2.353 | -0.092 | 8.493 | 1.017 | 1.144 |
| 140 | 6.159 | 122.314 | 124.135 | 1.015 | 2.087 | 0.006 | 7.787 | 1.042 | 1.317 |
| 141 | 3.699 | 26.491 | 50.013 | 1.888 | 1.423 | 0.276 | 6.082 | 1.164 | 1.914 |
| 142 | 2.911 | 12.912 | 30.993 | 2.400 | 1.111 | 0.380 | 4.602 | 1.164 | 1.841 |
| 143 | 3.260 | 18.141 | 38.324 | 2.113 | 1.259 | 0.325 | 5.083 | 1.148 | 1.790 |
| 144 | 4.718 | 54.990 | 74.190 | 1.349 | 1.740 | 0.130 | 6.404 | 1.061 | 1.440 |
| 145 | 4.222 | 39.403 | 63.365 | 1.608 | 1.596 | 0.206 | 6.307 | 1.132 | 1.690 |
| 146 | 4.383 | 44.077 | 66.467 | 1.508 | 1.644 | 0.178 | 5.695 | 1.101 | 1.431 |
| 147 | 4.592 | 50.702 | 72.311 | 1.426 | 1.705 | 0.154 | 6.339 | 1.092 | 1.507 |
| 148 | 5.025 | 66.423 | 83.552 | 1.258 | 1.822 | 0.100 | 6.336 | 1.053 | 1.328 |
| 149 | 4.963 | 64.001 | 82.663 | 1.292 | 1.806 | 0.111 | 6.499 | 1.068 | 1.399 |
| 150 | 4.599 | 50.934 | 69.989 | 1.374 | 1.707 | 0.138 | 6.146 | 1.053 | 1.408 |
| 151 | 2.399 | 7.230 | 22.125 | 3.060 | 0.859 | 0.486 | 4.536 | 1.224 | 2.313 |
| 152 | 2.935 | 13.243 | 32.024 | 2.418 | 1.122 | 0.384 | 4.861 | 1.183 | 1.959 |
| 153 | 3.863 | 30.188 | 52.094 | 1.726 | 1.480 | 0.237 | 6.082 | 1.111 | 1.749 |
| 154 | 3.776 | 28.180 | 49.453 | 1.755 | 1.450 | 0.244 | 5.408 | 1.104 | 1.582 |
| 155 | 4.260 | 40.472 | 61.487 | 1.519 | 1.607 | 0.182 | 5.762 | 1.079 | 1.459 |
| 156 | 5.431 | 83.874 | 97.987 | 1.168 | 1.924 | 0.068 | 7.205 | 1.057 | 1.403 |
| 157 | 4.765 | 56.635 | 76.086 | 1.343 | 1.753 | 0.128 | 6.498 | 1.067 | 1.455 |
| 158 | 5.039 | 66.989 | 84.967 | 1.268 | 1.826 | 0.103 | 6.952 | 1.065 | 1.470 |
| 159 | 5.176 | 72.588 | 88.778 | 1.223 | 1.861 | 0.087 | 6.919 | 1.055 | 1.410 |
| 160 | 4.666 | 53.201 | 74.190 | 1.395 | 1.726 | 0.144 | 6.855 | 1.085 | 1.593 |
| 161 | 3.064 | 15.063 | 33.363 | 2.215 | 1.178 | 0.345 | 5.407 | 1.131 | 1.996 |
| 162 | 2.089 | 4.773 | 16.676 | 3.494 | 0.679 | 0.543 | 3.802 | 1.216 | 2.214 |
| 163 | 4.066 | 35.201 | 57.898 | 1.645 | 1.547 | 0.216 | 6.408 | 1.115 | 1.757 |
| 164 | 4.377 | 43.895 | 65.607 | 1.495 | 1.642 | 0.175 | 6.497 | 1.090 | 1.618 |
| 165 | 4.039 | 34.501 | 55.010 | 1.594 | 1.538 | 0.203 | 5.729 | 1.073 | 1.522 |
| 166 | 4.481 | 47.109 | 68.489 | 1.454 | 1.673 | 0.163 | 6.402 | 1.086 | 1.551 |
| 167 | 4.209 | 39.047 | 59.740 | 1.530 | 1.592 | 0.185 | 5.826 | 1.073 | 1.486 |
| 168 | 3.822 | 29.239 | 49.233 | 1.684 | 1.466 | 0.226 | 5.405 | 1.073 | 1.517 |
| 169 | 4.702 | 54.446 | 74.589 | 1.370 | 1.736 | 0.137 | 6.697 | 1.074 | 1.529 |
| 170 | 4.192 | 38.581 | 59.936 | 1.554 | 1.586 | 0.191 | 6.146 | 1.085 | 1.591 |
| 171 | 2.160 | 5.276 | 17.737 | 3.362 | 0.722 | 0.527 | 4.116 | 1.210 | 2.306 |
| 172 | 1.705 | 2.597 | 11.555 | 4.450 | 0.414 | 0.648 | 3.574 | 1.265 | 2.651 |
| 173 | 2.583 | 9.023 | 24.742 | 2.742 | 0.955 | 0.438 | 4.668 | 1.181 | 2.133 |
| 174 | 3.126 | 16.001 | 35.081 | 2.192 | 1.204 | 0.341 | 5.246 | 1.142 | 1.917 |
| 175 | 2.525 | 8.434 | 23.937 | 2.838 | 0.926 | 0.453 | 4.444 | 1.195 | 2.102 |
| 176 | 2.312 | 6.469 | 19.989 | 3.090 | 0.811 | 0.490 | 4.055 | 1.191 | 2.088 |
| 177 | 3.109 | 15.734 | 33.572 | 2.134 | 1.197 | 0.329 | 4.796 | 1.106 | 1.706 |
| 178 | 3.436 | 21.237 | 40.904 | 1.926 | 1.327 | 0.285 | 5.117 | 1.103 | 1.643 |
| 179 | 3.966 | 32.661 | 53.479 | 1.637 | 1.514 | 0.214 | 5.762 | 1.082 | 1.572 |
| 180 | 2.914 | 12.959 | 30.727 | 2.371 | 1.113 | 0.375 | 5.182 | 1.152 | 2.048 |
| 181 | 1.549 | 1.947 | 10.031 | 5.151 | 0.289 | 0.712 | 3.255 | 1.330 | 2.794 |
| 182 | 1.884 | 3.501 | 13.132 | 3.751 | 0.544 | 0.574 | 4.057 | 1.178 | 2.537 |
| 183 | 1.659 | 2.389 | 11.202 | 4.689 | 0.378 | 0.671 | 3.349 | 1.296 | 2.618 |
| 184 | 3.007 | 14.242 | 32.749 | 2.299 | 1.154 | 0.362 | 5.147 | 1.153 | 1.972 |

| ID | D (mm) | V (mm³) | A (mm²) | A/V | log(V) | log(A/V) | L (mm) | $S^{-1}$ | M |
|---|---|---|---|---|---|---|---|---|---|
| 185 | 1.936 | 3.797 | 15.359 | 4.045 | 0.579 | 0.607 | 3.963 | 1.305 | 2.672 |
| 186 | 2.684 | 10.122 | 26.494 | 2.618 | 1.005 | 0.418 | 4.862 | 1.171 | 2.121 |
| 187 | 3.088 | 15.421 | 34.598 | 2.244 | 1.188 | 0.351 | 5.379 | 1.155 | 2.011 |
| 188 | 2.792 | 11.399 | 28.206 | 2.474 | 1.057 | 0.393 | 4.827 | 1.152 | 1.991 |
| 189 | 2.513 | 8.308 | 22.889 | 2.755 | 0.919 | 0.440 | 4.377 | 1.154 | 2.010 |
| 190 | 2.294 | 6.324 | 19.457 | 3.077 | 0.801 | 0.488 | 4.314 | 1.176 | 2.212 |
| 191 | 2.095 | 4.814 | 16.292 | 3.384 | 0.683 | 0.529 | 4.090 | 1.182 | 2.307 |
| 192 | 2.056 | 4.550 | 16.156 | 3.551 | 0.658 | 0.550 | 4.379 | 1.217 | 2.591 |
| 193 | 1.454 | 1.609 | 8.334 | 5.180 | 0.206 | 0.714 | 3.125 | 1.255 | 2.698 |
| 194 | 1.383 | 1.384 | 7.839 | 5.663 | 0.141 | 0.753 | 3.252 | 1.305 | 3.069 |
| 195 | 1.940 | 3.825 | 14.705 | 3.845 | 0.583 | 0.585 | 4.028 | 1.243 | 2.581 |
| 196 | 1.658 | 2.386 | 10.920 | 4.576 | 0.378 | 0.661 | 3.607 | 1.265 | 2.751 |
| 197 | 2.270 | 6.128 | 20.030 | 3.269 | 0.787 | 0.514 | 4.695 | 1.237 | 2.558 |
| 198 | 1.936 | 3.797 | 14.148 | 3.726 | 0.579 | 0.571 | 3.738 | 1.202 | 2.321 |
| 199 | 2.327 | 6.596 | 19.988 | 3.030 | 0.819 | 0.481 | 4.248 | 1.175 | 2.145 |
| 200 | 2.566 | 8.845 | 23.615 | 2.670 | 0.947 | 0.426 | 4.571 | 1.142 | 2.034 |

Table A.7. Geometric properties of 60 Florida limestone particles

| ID | D (mm) | V (mm³) | A (mm²) | A/V | log(V) | log(A/V) | L (mm) | $S^{-1}$ | M |
|---|---|---|---|---|---|---|---|---|---|
| 1 | 6.134 | 120.867 | 160.352 | 1.327 | 2.082 | 0.123 | 10.283 | 1.356 | 2.274 |
| 2 | 5.032 | 66.718 | 103.106 | 1.545 | 1.824 | 0.189 | 7.319 | 1.296 | 1.885 |
| 3 | 5.043 | 67.160 | 99.098 | 1.476 | 1.827 | 0.169 | 7.065 | 1.240 | 1.737 |
| 4 | 4.713 | 54.805 | 91.780 | 1.675 | 1.739 | 0.224 | 7.471 | 1.315 | 2.085 |
| 5 | 6.236 | 126.957 | 168.909 | 1.330 | 2.104 | 0.124 | 9.923 | 1.383 | 2.200 |
| 6 | 7.372 | 209.745 | 210.451 | 1.003 | 2.322 | 0.001 | 10.694 | 1.233 | 1.788 |
| 7 | 5.177 | 72.655 | 114.487 | 1.576 | 1.861 | 0.197 | 8.310 | 1.360 | 2.182 |
| 8 | 4.156 | 37.583 | 70.584 | 1.878 | 1.575 | 0.274 | 6.872 | 1.301 | 2.151 |
| 9 | 4.026 | 34.176 | 70.121 | 2.052 | 1.534 | 0.312 | 8.405 | 1.377 | 2.874 |
| 10 | 5.588 | 91.345 | 121.589 | 1.331 | 1.961 | 0.124 | 8.572 | 1.240 | 1.902 |
| 11 | 5.126 | 70.537 | 97.418 | 1.381 | 1.848 | 0.140 | 7.057 | 1.180 | 1.624 |
| 12 | 4.850 | 59.720 | 90.196 | 1.510 | 1.776 | 0.179 | 7.487 | 1.221 | 1.885 |
| 13 | 3.631 | 25.067 | 50.839 | 2.028 | 1.399 | 0.307 | 5.309 | 1.227 | 1.794 |
| 14 | 5.365 | 80.855 | 107.984 | 1.336 | 1.908 | 0.126 | 8.241 | 1.194 | 1.834 |
| 15 | 4.929 | 62.684 | 98.103 | 1.565 | 1.797 | 0.195 | 7.630 | 1.286 | 1.990 |
| 16 | 4.626 | 51.841 | 80.691 | 1.557 | 1.715 | 0.192 | 5.771 | 1.200 | 1.497 |
| 17 | 5.029 | 66.591 | 95.346 | 1.432 | 1.823 | 0.156 | 7.559 | 1.200 | 1.804 |
| 18 | 5.535 | 88.808 | 116.817 | 1.315 | 1.948 | 0.119 | 7.316 | 1.214 | 1.604 |
| 19 | 4.660 | 53.001 | 83.361 | 1.573 | 1.724 | 0.197 | 6.529 | 1.222 | 1.711 |
| 20 | 3.915 | 31.422 | 58.956 | 1.876 | 1.497 | 0.273 | 5.794 | 1.224 | 1.812 |
| 21 | 3.795 | 28.623 | 55.863 | 1.952 | 1.457 | 0.290 | 5.244 | 1.235 | 1.706 |
| 22 | 3.027 | 14.524 | 37.537 | 2.585 | 1.162 | 0.412 | 4.912 | 1.304 | 2.116 |

| | | | | | | | | |
|---|---|---|---|---|---|---|---|---|
| 23 | 4.524 | 48.494 | 83.481 | 1.721 | 1.686 | 0.236 | 7.444 | 1.298 | 2.136 |
| 24 | 5.367 | 80.956 | 110.756 | 1.368 | 1.908 | 0.136 | 7.650 | 1.224 | 1.744 |
| 25 | 3.569 | 23.812 | 49.738 | 2.089 | 1.377 | 0.320 | 6.038 | 1.243 | 2.102 |
| 26 | 3.665 | 25.768 | 53.536 | 2.078 | 1.411 | 0.318 | 5.368 | 1.269 | 1.859 |
| 27 | 4.380 | 43.983 | 75.620 | 1.719 | 1.643 | 0.235 | 6.449 | 1.255 | 1.848 |
| 28 | 4.484 | 47.200 | 77.676 | 1.646 | 1.674 | 0.216 | 7.204 | 1.230 | 1.976 |
| 29 | 6.540 | 146.444 | 166.775 | 1.139 | 2.166 | 0.056 | 9.741 | 1.241 | 1.849 |
| 30 | 5.456 | 85.058 | 117.265 | 1.379 | 1.930 | 0.139 | 8.183 | 1.254 | 1.880 |
| 31 | 6.014 | 113.910 | 141.973 | 1.246 | 2.057 | 0.096 | 9.544 | 1.249 | 1.983 |
| 32 | 5.549 | 89.465 | 140.364 | 1.569 | 1.952 | 0.196 | 9.732 | 1.451 | 2.545 |
| 33 | 4.785 | 57.360 | 90.697 | 1.581 | 1.759 | 0.199 | 7.703 | 1.261 | 2.030 |
| 34 | 6.119 | 119.983 | 141.748 | 1.181 | 2.079 | 0.072 | 9.030 | 1.205 | 1.778 |
| 35 | 4.313 | 41.996 | 75.571 | 1.799 | 1.623 | 0.255 | 7.369 | 1.293 | 2.210 |
| 36 | 4.433 | 45.614 | 80.988 | 1.776 | 1.659 | 0.249 | 6.766 | 1.312 | 2.002 |
| 37 | 5.119 | 70.254 | 114.845 | 1.635 | 1.847 | 0.213 | 9.250 | 1.395 | 2.520 |
| 38 | 6.741 | 160.402 | 176.175 | 1.098 | 2.205 | 0.041 | 9.355 | 1.234 | 1.712 |
| 39 | 6.416 | 138.308 | 160.196 | 1.158 | 2.141 | 0.064 | 9.277 | 1.239 | 1.791 |
| 40 | 6.043 | 115.545 | 156.401 | 1.354 | 2.063 | 0.131 | 9.588 | 1.363 | 2.163 |
| 41 | 5.908 | 107.950 | 135.627 | 1.256 | 2.033 | 0.099 | 8.171 | 1.237 | 1.711 |
| 42 | 6.096 | 118.594 | 149.433 | 1.260 | 2.074 | 0.100 | 8.887 | 1.280 | 1.866 |
| 43 | 5.549 | 89.478 | 119.948 | 1.341 | 1.952 | 0.127 | 8.520 | 1.240 | 1.904 |
| 44 | 5.779 | 101.033 | 133.760 | 1.324 | 2.004 | 0.122 | 8.963 | 1.275 | 1.978 |
| 45 | 5.434 | 83.993 | 114.994 | 1.369 | 1.924 | 0.136 | 7.547 | 1.240 | 1.722 |
| 46 | 5.227 | 74.775 | 110.463 | 1.477 | 1.874 | 0.169 | 8.980 | 1.287 | 2.211 |
| 47 | 5.737 | 98.881 | 135.019 | 1.365 | 1.995 | 0.135 | 8.747 | 1.306 | 1.991 |
| 48 | 4.758 | 56.413 | 81.732 | 1.449 | 1.751 | 0.161 | 6.118 | 1.149 | 1.477 |
| 49 | 4.478 | 47.006 | 77.063 | 1.639 | 1.672 | 0.215 | 6.500 | 1.223 | 1.776 |
| 50 | 5.722 | 98.099 | 119.932 | 1.223 | 1.992 | 0.087 | 8.531 | 1.166 | 1.738 |
| 51 | 5.189 | 73.159 | 109.218 | 1.493 | 1.864 | 0.174 | 7.577 | 1.291 | 1.885 |
| 52 | 5.940 | 109.741 | 134.225 | 1.223 | 2.040 | 0.087 | 8.293 | 1.211 | 1.691 |
| 53 | 6.812 | 165.476 | 176.663 | 1.068 | 2.219 | 0.028 | 10.125 | 1.212 | 1.802 |
| 54 | 4.808 | 58.196 | 93.870 | 1.613 | 1.765 | 0.208 | 7.107 | 1.293 | 1.910 |
| 55 | 5.372 | 81.168 | 118.436 | 1.459 | 1.909 | 0.164 | 8.202 | 1.306 | 1.995 |
| 56 | 7.048 | 183.353 | 206.774 | 1.128 | 2.263 | 0.052 | 11.983 | 1.325 | 2.252 |
| 57 | 5.529 | 88.478 | 127.732 | 1.444 | 1.947 | 0.159 | 8.666 | 1.330 | 2.085 |
| 58 | 4.354 | 43.219 | 75.688 | 1.751 | 1.636 | 0.243 | 6.662 | 1.271 | 1.945 |
| 59 | 5.342 | 79.817 | 107.018 | 1.341 | 1.902 | 0.127 | 8.036 | 1.194 | 1.796 |
| 60 | 5.042 | 67.102 | 93.675 | 1.396 | 1.827 | 0.145 | 6.214 | 1.173 | 1.446 |

# REFERENCES

1. Su, Y.F., Bhattacharya, S., Lee, S.J., Lee, C.H., Shin, M.: A new interpretation of three-dimensional particle geometry: M-A-V-L. Transp. Geotech. 23, 100328 (2020). https://doi.org/10.1016/j.trgeo.2020.100328

2. Yang, J., Luo, X.D.: Exploring the relationship between critical state and particle shape for granular materials. J. Mech. Phys. Solids. 84, 196–213 (2015). https://doi.org/10.1016/j.jmps.2015.08.001

3. Ashmawy, A.K., Sukumaran, B., Hoang, V.V.: Evaluating The Influence of Particle Shape on Liquefaction Behavior Using Discrete Element Modeling. In: Chung, J.S. and Prinsenberg, S. (eds.) Proc. 13th Int.Offshore and Polar Engeering Conf. pp. 542–549. International Society of Offshore and Polar Engineers, Honolulu, Hawaii, USA (2003)

4. Terzaghi, K., Peck, R.B., Mesri, G.: Soil mechanics in engineering practice. Wiley (1996)

5. Mitchell, J.K., Soga, K.: Fundamentals of soil behavior. John Wiley & Sons (2005)

6. Cho, G.-C., Dodds, J., Santamarina, J.C.: Particle Shape Effects on Packing Density, Stiffness, and Strength: Natural and Crushed Sands. J. Geotech. Geoenvironmental Eng. 132, 591–602 (2006). https://doi.org/10.1061/(ASCE)1090-0241(2006)132:5(591)

7. Sutherland, C.A.M., Liu, X., Zhang, L., Chu, Y., Oldmeadow, J.A., Young, A.W.: Facial First Impressions Across Culture: Data-Driven Modeling of Chinese and British Perceivers' Unconstrained Facial Impressions. Personal. Soc. Psychol. Bull. 44, 521–537 (2018). https://doi.org/10.1177/0146167217744194

8. DeBruine, L., Jones, B.: Face Research, http://faceresearch.org/

9. Lee, S.J., Lee, C.H., Shin, M., Bhattacharya, S., Su, Y.F.: Influence of coarse aggregate angularity on the mechanical performance of cement-based materials. Constr. Build. Mater. 204, 184–192 (2019). https://doi.org/10.1016/j.conbuildmat.2019.01.135

10. Bhattacharya, S., Subedi, S., Lee, S.J., Pradhananga, N.: Estimation of 3D Sphericity by Volume Measurement – Application to Coarse Aggregates. Transp. Geotech. 23, 100344 (2020). https://doi.org/10.1016/j.trgeo.2020.100344

11. Polyga: Polyga Compact C504, https://www.polyga.com/products/compact-c504/

12. Zheng, W., Hu, X., Tannant, D.D., Zhang, K., Xu, C.: Characterization of two- and three-dimensional morphological properties of fragmented sand grains. Eng. Geol. 263, 105358 (2019). https://doi.org/10.1016/j.enggeo.2019.105358

13. Wadell, H.: Sphericity and Roundness of Rock Particles. J. Geol. 41, 310–331 (1933)

14. Barrett, P.J.: The shape of rock particles, a critical review. Sedimentology. 27, 291–303 (1980). https://doi.org/10.1111/j.1365-3091.1980.tb01179.x

15. Zhao, B., Wang, J.: 3D quantitative shape analysis on form, roundness, and compactness

with µCT. Powder Technol. 291, 262–275 (2016). https://doi.org/10.1016/j.powtec.2015.12.029

16. Jerves, A.X., Kawamoto, R.Y., Andrade, J.E.: Effects of grain morphology on critical state: a computational analysis. Acta Geotech. 11, 493–503 (2016). https://doi.org/10.1007/s11440-015-0422-8

17. Richard, P., Nicodemi, M., Delannay, R., Ribière, P., Bideau, D.: Slow relaxation and compaction of granular systems. Nat. Mater. 4, 121–128 (2005). https://doi.org/10.1038/nmat1300